\documentclass[times,tighten,twocolumn,twocolappendix,resetfootnote]{aastex701}

\usepackage{graphicx,color}
\usepackage{url}

\newcommand{\xmm}{{\em XMM-Newton}}

\newcommand{\swift}{{\em Swift}}

\newcommand{\nicer}{NICER}
\newcommand{\nustar}{{\em NuSTAR}}

\newcommand{\ros}{ROSAT}
\newcommand{\gaia}{{\em Gaia}}
\newcommand{\srcfirst}{\object{EP\,J174942.2$-$384834}}
\newcommand{\src}{EP\,J1749}

\newcommand{\be}{\begin{equation}}
\newcommand{\en}{\end{equation}}

\def\arcmin{\mbox{$^{\prime}$}}

\def\nh{\hbox{$N_{\rm H}$}}
\def\flux {\mbox{erg~cm$^{-2}$~s$^{-1}$}}
\def\lum {\mbox{erg~s$^{-1}$}}

\shorttitle{\src: a new very faint X-ray transient}
\shortauthors{Coti~Zelati et al.}
\received{2026 January 23}
\accepted{2026 April 28}
\submitjournal{ApJ}

\begin{document}

\title{Multi-wavelength outburst activity from \srcfirst: \\a very faint X-ray transient discovered by Einstein Probe}

\correspondingauthor{F.~Coti~Zelati}
\author[0000-0001-7611-1581]{F.~Coti~Zelati}
\affiliation{Institute of Space Sciences (ICE, CSIC), Campus UAB, Carrer de Can Magrans s/n, Barcelona, E-08193, Spain}
\affiliation{Institut d’Estudis Espacials de Catalunya (IEEC), Barcelona, E-08034, Spain}
\affiliation{INAF--Osservatorio Astronomico di Brera, Via Bianchi 46, I-23807 Merate (LC), Italy}
\email[show]{cotizelati@ice.csic.es}

\author[0000-0001-5674-4664]{A.~Marino}
\affiliation{Institute of Space Sciences (ICE, CSIC), Campus UAB, Carrer de Can Magrans s/n, Barcelona, E-08193, Spain}
\affiliation{Institut d’Estudis Espacials de Catalunya (IEEC), Barcelona, E-08034, Spain}
\affiliation{INAF, Istituto di Astrofisica Spaziale e Fisica Cosmica, Via U. La Malfa 153, I-90146 Palermo, Italy}
\email[]{marino@ice.csic.es}

\author[0009-0009-1721-3663]{Y.~L.~Wang}
\affiliation{National Astronomical Observatories, Chinese Academy of Sciences, 20A Datun Road, Beijing 100101, China}
\affiliation{Institute of Space Sciences (ICE, CSIC), Campus UAB, Carrer de Can Magrans s/n, Barcelona, E-08193, Spain}
\affiliation{Institut d’Estudis Espacials de Catalunya (IEEC), Barcelona, E-08034, Spain}
\affiliation{School of Astronomy and Space Science, University of Chinese Academy of Sciences, 19A Yuquan Road, Beijing 100049, China}
\email[]{wangyilong@nao.cas.cn}

\author[0000-0002-0146-3096]{M.~Veresvarska}
\affiliation{Institute of Space Sciences (ICE, CSIC), Campus UAB, Carrer de Can Magrans s/n, Barcelona, E-08193, Spain}
\affiliation{Institut d’Estudis Espacials de Catalunya (IEEC), Barcelona, E-08034, Spain}
\email[]{mveresvarska@ice.csic.es}

\author[0000-0003-2177-6388]{N.~Rea}
\affiliation{Institute of Space Sciences (ICE, CSIC), Campus UAB, Carrer de Can Magrans s/n, Barcelona, E-08193, Spain}
\affiliation{Institut d’Estudis Espacials de Catalunya (IEEC), Barcelona, E-08034, Spain}
\email[]{rea@ice.csic.es}

\author[0000-0002-6449-106X]{S.~Guillot}
\affiliation{Institut de Recherche en Astrophysique et Plan\'{e}tologie, UPS-OMP, CNRS, CNES, 9 avenue du Colonel Roche, BP 44346, Toulouse Cedex 4, 31028, France}
\email[]{sebastien.guillot@irap.omp.eu}

\author[0000-0002-7004-9956]{D.~A.~H.~Buckley}
\affiliation{South African Astronomical Observatory, P.O Box 9, Observatory, 7935 Cape Town, South Africa}
\affiliation{Department of Astronomy, University of Cape Town, Private Bag X3, Rondebosch 7701, South Africa}
\affiliation{Department of Physics, University of the Free State, PO Box 339, Bloemfontein 9300, South Africa}
\email[]{DAH.Buckley@saao.nrf.ac.za}

\author[0000-0002-4633-6832]{N.~Rawat}
\affiliation{South African Astronomical Observatory, P.O Box 9, Observatory, 7935 Cape Town, South Africa}
\email[]{rawatn@saao.ac.za}

\author[0000-0002-6154-5843]{S.~E.~Motta}
\affiliation{INAF--Osservatorio Astronomico di Brera, Via Bianchi 46, I-23807 Merate (LC), Italy}
\email[]{sara.motta@inaf.it}

\author[0000-0003-2443-3698]{Y.~Xu}
\affiliation{Key Laboratory of Particle Astrophysics, Institute of High Energy Physics, Chinese Academy of Sciences, Beijing 100049, China}
\email[]{xuyj@ihep.ac.cn}

\author[0000-0003-2310-8105]{Z.~Li}
\affiliation{School of Science, Qingdao University of Technology, Qingdao 266525, China}
\email[]{lizhaosheng@xtu.edu.cn}

\author[0000-0001-7199-2906]{Y.-F.~Huang}
\affiliation{School of Astronomy and Space Science, Nanjing University, Nanjing 210023, China}
\affiliation{Key Laboratory of Modern Astronomy and Astrophysics (Nanjing University), Ministry of Education, China}
\email[]{hyf@nju.edu.cn}

\author[0000-0001-7584-6236]{H.~Feng}
\affiliation{Key Laboratory of Particle Astrophysics, Institute of High Energy Physics, Chinese Academy of Sciences, Beijing 100049, China}
\email[]{hfeng@ihep.ac.cn}

\author[0000-0002-2705-4338]{L.~Tao}
\affiliation{Key Laboratory of Particle Astrophysics, Institute of High Energy Physics, Chinese Academy of Sciences, Beijing 100049, China}
\email[]{taolian@ihep.ac.cn}

\author[0000-0001-8688-9784]{M.~Imbrogno}
\affiliation{Institute of Space Sciences (ICE, CSIC), Campus UAB, Carrer de Can Magrans s/n, Barcelona, E-08193, Spain}
\affiliation{Institut d’Estudis Espacials de Catalunya (IEEC), Barcelona, E-08034, Spain}
\affiliation{INAF -- Osservatorio Astronomico di Roma, Via Frascati 33, I-00078, Monte Porzio Catone (RM), Italy}
\email[]{mimbrogno@ice.csic.es}

\author[0000-0003-4795-7072]{G.~Illiano}
\affiliation{Institute of Space Sciences (ICE, CSIC), Campus UAB, Carrer de Can Magrans s/n, Barcelona, E-08193, Spain}
\affiliation{Institut d’Estudis Espacials de Catalunya (IEEC), Barcelona, E-08034, Spain}
\affiliation{INAF--Osservatorio Astronomico di Brera, Via Bianchi 46, I-23807 Merate (LC), Italy}
\email[]{giulia.illiano@inaf.it}

\author[0000-0003-1285-4057]{M.~C.~Baglio}
\affiliation{INAF--Osservatorio Astronomico di Brera, Via Bianchi 46, I-23807 Merate (LC), Italy}
\email[]{maria.baglio@inaf.it}

\author[0000-0003-4200-9954]{H.~Q.~Cheng}
\affiliation{National Astronomical Observatories, Chinese Academy of Sciences, 20A Datun Road, Beijing 100101, China}
\email[]{hqcheng@bao.ac.cn}

\author[0000-0002-2006-1615]{C.~C.~Jin}
\affiliation{National Astronomical Observatories, Chinese Academy of Sciences, 20A Datun Road, Beijing 100101, China}
\affiliation{School of Astronomy and Space Science, University of Chinese Academy of Sciences, 19A Yuquan Road, Beijing 100049, China}
\affiliation{Institute for Frontier in Astronomy and Astrophysics, Beijing Normal University, Beijing 102206, China}
\email[]{ccjin@nao.cas.cn}

\author[0000-0002-9615-1481]{H.~Sun}
\affiliation{National Astronomical Observatories, Chinese Academy of Sciences, 20A Datun Road, Beijing 100101, China}
\email[]{hsun@nao.cas.cn}

\author{W.~Yuan}
\affiliation{National Astronomical Observatories, Chinese Academy of Sciences, 20A Datun Road, Beijing 100101, China}
\affiliation{School of Astronomy and Space Science, University of Chinese Academy of Sciences, 19A Yuquan Road, Beijing 100049, China}
\email[]{wmy@nao.cas.cn}

\author[0000-0002-0426-3276]{F.~Carotenuto}
\affiliation{INAF -- Osservatorio Astronomico di Roma, Via Frascati 33, I-00078, Monte Porzio Catone (RM), Italy}
\email[]{francesco.carotenuto@inaf.it}

\author[0000-0002-5654-2744]{R.~P.~Fender}
\affiliation{Department of Physics, University of Oxford, Denys Wilkinson Building, Keble Road, Oxford OX1 3RH, UK}
\email[]{rob.fender@physics.ox.ac.uk}

\author[0000-0003-0860-440X]{A.~Coleiro}
\affiliation{Université Paris Cité, CNRS, Astroparticule et Cosmologie, F-75013 Paris, France}
\email{coleiro@apc.in2p3.fr}

\author[0000-0001-9494-0981]{D.~G\"otz}
\affiliation{Université Paris-Saclay, Université Paris Cité, CEA, CNRS, AIM, 91191 Gif-sur-Yvette, France}
\email{diego.gotz@cea.fr}

\author{H.~L.~Li}
\affiliation{National Astronomical Observatories, Chinese Academy of Sciences, Beijing 100101, China}
\email{lhl@nao.cas.cn}

\author[0000-0001-5612-5185]{P.~Maggi}
\affiliation{Observatoire Astronomique de Strasbourg, Université de Strasbourg, CNRS, 11 rue de l’Université, F-67000 Strasbourg, France}
\email{pierre.maggi@astro.unistra.fr}

\author{Y.~L.~Qiu}
\affiliation{National Astronomical Observatories, Chinese Academy of Sciences, Beijing 100101, China}
\email{qiuyl@nao.cas.cn}

\author{J.~Wang}
\affiliation{National Astronomical Observatories, Chinese Academy of Sciences, Beijing 100101, China}
\email{wj@bao.ac.cn}

\author[0000-0002-9422-3437]{L.~P.~Xin}
\affiliation{National Astronomical Observatories, Chinese Academy of Sciences, Beijing 100101, China}
\email{xlp@nao.cas.cn}


\begin{abstract}
We report the discovery and multi-wavelength characterization of the Galactic transient \srcfirst, first detected by the \emph{Einstein Probe} during a faint X-ray outburst in March 2025. Coordinated follow-up observations revealed two major outbursts and a rebrightening over a seven-month period. Broadband X-ray spectral modeling shows that the outburst emission was dominated by thermal Comptonization of very soft seed photons. The absence of a detected thermal disk component, together with the low inferred seed-photon temperature, is consistent with a cool and possibly truncated accretion disk. The X-ray spectrum remained consistently hard throughout the outburst activity, with a power-law photon index of $\Gamma \approx 1$--2, gradually softening as the flux declined. The optical/UV counterpart brightened in tandem with the X-ray emission and exhibited a blue continuum with broad Balmer absorption features. Together with the optical/UV--X-ray luminosity correlation, this supports a disk-dominated origin of the optical/UV outburst emission, with viscous heating likely playing a major role and irradiation possibly contributing, especially in the UV. No radio counterpart was detected, implying at most very faint jet activity. Taken together, the observed properties support the classification of \srcfirst\ as a very faint X-ray transient black hole candidate. This study demonstrates the ability of \emph{Einstein Probe} to uncover and characterize the faintest accreting compact objects in the Galaxy.
\end{abstract}



\section{Introduction}
\label{sec:intro}

Low-mass X-ray binaries (LMXBs) hosting either neutron stars (NSs) or stellar-mass black holes (BHs) are prime laboratories for studying accretion physics, relativistic outflows, and the evolution of compact objects (for recent reviews, see \citealt{Kalemci2022,Bahramian2023,DiSalvo2023}; for catalogs, see \citealt{Avakyan2023,Armas-Padilla2023,Fortin2024}). In these systems, matter transferred from a low-mass companion star (typically with a mass $\lesssim 1$\,$\mathrm{M_\odot}$) through Roche-lobe overflow forms an accretion disk as it spirals inward toward the compact object \citep{Shakura1973}. Most LMXBs are transient: they remain in a quiescent state with X-ray luminosities of $L_{\rm X} \approx 10^{30}-10^{34}$\,\lum, punctuated by outbursts during which their luminosity increases by several orders of magnitude to $L_{\rm X} \approx 10^{36}$--$10^{38}$\,\lum, lasting from weeks to years \citep[e.g.,][]{Corral-Santana2016,Tetarenko2016,Heinke2025}. These outbursts are generally well explained by the disk instability model (DIM), in which thermal-viscous instabilities trigger episodic accretion \citep{Lasota2001,Dubus2001,Hameury2020}.

Over the past two decades, a subpopulation of very faint X-ray transients (VFXTs) has emerged, with outbursts peaking at $L_{\rm X}\lesssim 10^{36}$\,\lum\ \citep[e.g.,][]{Sakano2005,Wijnands2006,DelSanto2007,Degenaar2009,Sidoli2011,ArmasPadilla2013,Koch2014,Degenaar2015,Heinke2015,Bahramian2021,Stoop2021,Ahmed2024,Stel2026}. Most VFXTs are NS systems, as revealed by type-I X-ray bursts or coherent pulsations at the NS spin period, although a few BH candidates have shown similarly faint outbursts. These sources are particularly interesting because they probe accretion in a regime where radiative efficiency is extremely low, with much of the energy of the inflowing material advected onto the compact object or carried away in outflows. Their duty cycles, recurrence times, and population demographics remain poorly constrained, largely because their intrinsic faintness places them below the sensitivity thresholds of all-sky X-ray monitors. Indeed, most VFXTs have been identified serendipitously through deep, dedicated monitoring of the Galactic Center with sensitive pointed observations (e.g., \citealt{Degenaar2012,Degenaar2015}).

VFXTs display distinctive multi-wavelength signatures (e.g., \citealt{ArmasPadilla2013, Weng2015}). Their X-ray emission is typically well described by a power-law spectrum with photon index $\Gamma \approx 1$--$2$, consistent with Compton upscattering of photons from a truncated disk by an optically thin coronal flow\footnote{Because VFXTs are intrinsically faint, any soft thermal component is only detectable when observations of sufficiently high quality are available (e.g., \citealt{Armas-Padilla2014a}).}. Many sources remain in this hard state throughout the outburst and exhibit gradual spectral softening during the decay phase (e.g., \citealt{Wijnands2015}). In BH systems, this behavior is commonly interpreted as the signature of the inner flow transitioning into a radiatively inefficient state as the disk progressively recedes and is replaced by a hot, geometrically thick flow (e.g., \citealt{Narayan1994,Narayan1995,Menou2000,Dubus2001}).
The optical and UV emission in outburst is produced primarily in the outer disk and is often dominated by viscous heating (e.g., \citealt{ArmasPadilla2013,Weng2015}). In some cases, compact jets may contribute non-thermal emission up to the near-infrared (NIR; e.g. \citealt{Plotkin2016,Russell2018}). Radio detections remain rare: a few BH VFXTs exhibit flat or slightly inverted spectra at low X-ray luminosities, consistent with optically thick, partially self-absorbed synchrotron emission from a steady compact jet powered by continuous particle acceleration \citep[e.g.,][]{Sivakoff2011,Plotkin2016,Stoop2021}. In contrast, NS VFXTs are systematically more radio‑faint \citep{vandenEijnden2021}.

The prospects of detecting and characterizing VFXTs are now entering a new era after the advent of the \emph{Einstein Probe} (EP; \citealt{Yuan2022,Yuan2025}) mission. Thanks to its wide-field X-ray imaging and high-cadence all-sky coverage, EP can detect X-ray transients that were previously beyond reach, enabling detailed studies of the faintest accreting binaries in our Galaxy. This capability was already demonstrated by the recent discovery of a transient LMXB which underwent a $\simeq$20-day faint outburst with multi-wavelength properties consistent with a BH accretor \citep{Cheng2025b}. This event highlights the potential of EP to uncover an as-yet-unseen population of VFXTs.

This paper reports on the discovery and multi-wavelength study of \srcfirst\ (also known as EP250305a; hereafter \src), a new VFXT first detected by EP in March 2025. The rest of the paper is organized as follows: Section\,\ref{sec:obs} details the observing campaign and the data reduction procedures; Section\,\ref{sec:analysis} presents the data analysis and results, covering the source localization, counterpart identification, distance estimation, the temporal evolution across all bands, X-ray timing and spectral properties during outburst, optical spectroscopy and radio emission properties, and constraints on pre-outburst activity. Section\,\ref{sec:discussion} discusses the multi-wavelength properties of \src\ in the context of LMXBs. Conclusions follow in Section\,\ref{sec:conclusions}.

\section{Observations and Data Processing}
\label{sec:obs}

\subsection{X-ray Observations}
We analyzed X-ray observations of \src\ obtained with multiple observatories between March and September 2025. A log of these data is provided in Table\,\ref{tab:obsX}. Below we summarize the reduction procedures and analysis choices adopted for each instrument.

\subsubsection{EP}
\label{sec:ep}
The Wide-field X-ray Telescope (WXT; \citealt{Cheng2025} and references therein) onboard EP first detected X-ray emission from \src\ on 2025 March 5, after a gap of $>$5 months since the previous observation of the field on 2024 September 29. Subsequent WXT monitoring yielded detections only four times between March 5 and March 24 and 15 times between June 8 and July 4, whereas the remaining observations provided $3\sigma$ upper limits on the observed flux of order $\sim10^{-11}$--$10^{-10}$\,\flux. Because the individual exposures were relatively short ($\simeq$1.0--6.2\,ks) and the detections were often marginal, we co-added the spectral files within each of these two intervals to derive more robust flux estimates through spectral fitting.

The Follow-up X-ray Telescope (FXT; \citealt{Chen2021,Chen2025,Zhang2025}) observed \src\ 11 times between April and July 2025, with both telescope units (FXT-A and FXT-B) operating in Full Frame mode (time resolution of 50\,ms). We processed the data using the \texttt{fxtchain} routine included in the FXT Data Analysis Software package \citep{Zhao2025}, together with the FXT Calibration Database v1.20.
Source photons were extracted from a circular region of radius 60\arcsec\ centered on the source position (see Section\,\ref{sec:position}), and the background was estimated from a concentric annulus with inner and outer radii of 80\arcsec\ and 160\arcsec. The pipeline produced cleaned event files, spectra, response files, and light curves. For non-detections, we computed 3$\sigma$ upper limits on the net count rate at the source position using the \texttt{sosta} tool in \texttt{ximage}, adopting the same extraction regions; the derived limits were not strongly sensitive to the exact choice of region size.

\subsubsection{\nicer}
\label{sec:nicer}
The \emph{Neutron star Interior Composition Explorer} (\nicer; \citealt{Gendreau2016}) observed \src\ four times between 2025 March 26 and April 17. We processed the data using the \texttt{nicerl2} pipeline and removed intervals affected by high-energy particle-induced flares from the cleaned event files with the \texttt{flaghighenergyflares} task in the \texttt{nicerutil} package\footnote{\url{https://github.com/georgeyounes/NICERUTIL}}. Spectral products and background-subtracted light curves were extracted with \texttt{nicerl3-spect} and \texttt{nicerl3-lc}, respectively, adopting the SCORPEON background model in both cases\footnote{\url{https://heasarc.gsfc.nasa.gov/docs/nicer/analysis_threads/scorpeon-overview/}}.
Spectra were grouped to contain a minimum of 20 counts per energy channel.

\subsubsection{\nustar}
\label{sec:nustar}
The \emph{Nuclear Spectroscopic Telescope Array} (\nustar; \citealt{Harrison2013}) observed \src\ on 2025 March 30--31, with a total elapsed time of $\simeq$38\,ks and an on-source exposure of $\simeq$21.5\,ks. We processed the data using \texttt{nupipeline} with standard settings. Source events were extracted from a 100\arcsec\ circular region centered on the source position, whereas background events were taken from a nearby source-free circular region of the same size. The source was significantly detected up to $\approx$40\,keV, so we adopted the 3--40\,keV range for all subsequent timing and spectral analyses. Background-subtracted spectra and light curves for both focal plane modules (FPMA and FPMB) were extracted with \texttt{nuproducts}, and the spectra were grouped to contain a minimum of 50 counts per energy channel.

\subsubsection{SVOM/MXT}
\label{sec:svom_x}
The Microchannel X-ray Telescope (MXT; \citealt{Gotz2026}), the soft X-ray (0.3--10\,keV) ``lobster-eye'' telescope on board the \emph{Space-based multi-band astronomical Variable Objects Monitor} (SVOM; \citealt{Cordier2026}), observed \src\ from 2025 March 30 at 22:42:15 to March 31 at 06:30:59 (UTC). After filtering out intervals affected by stray light, Earth occultations, and passages through the South Atlantic Anomaly, the net exposure time was $\simeq$8.4\,ks. The event-mode data were processed with the MXT pipeline \citep{Maggi2026}. Source detection was performed by fitting the MXT point spread function using single- to quadruple-pixel events in the 0.3--10\,keV band, yielding a source position of R.A. = 17$^\mathrm{h}$49$^\mathrm{m}$37$\fs$49, Decl. = --38$^{\circ}$48$^{\prime}$53$\farcs$7 (J2000.0), with a 90\% confidence statistical uncertainty of 45\arcsec.
Source and background spectra were extracted from the event list at that position. For spectral fitting, we used the ancillary response file (ARF) and redistribution matrix file (RMF) provided by the MXT Instrument Centre, after rescaling the 0.3--0.6\,keV ARF to account for the effective-area loss caused by bad pixels identified by the pipeline. Inspection of the MXT image and fitted PSF suggests that the observation is affected by non-X-ray background contamination at a level that is difficult to quantify. We therefore conservatively treat the MXT flux measurement as an upper limit.

\subsubsection{\swift/XRT}
\label{sec:swift_x}
The \emph{Neil Gehrels Swift Observatory} (\swift; \citealt{Gehrels2004}) observed \src\ on 29 occasions between 2025 March 30 and September 21, with exposure times typically ranging from fractions of a ks to a few ks. In all cases, the X-Ray Telescope (XRT; \citealt{Burrows2005}) operated in Photon Counting (PC) mode, providing a time resolution of 2.5\,s. We processed the data with the standard \texttt{xrtpipeline} task. The source was detected in most observations, always at net count rates $\lesssim 0.4$\,counts\,s$^{-1}$, so pile-up corrections were not required. For the epochs with detections, source photons were extracted from a circular region of radius $\simeq$47\arcsec\ centered on the source position, whereas background events were extracted from an annulus with inner and outer radii of $\simeq$94\arcsec\ and 188\arcsec, respectively. Background-subtracted spectra and light curves were obtained with \texttt{xselect}, ancillary response files were generated with \texttt{xrtmkarf}, and the appropriate response matrices were assigned from the calibration database.

For non-detections, we estimated 3$\sigma$ upper limits on the net count rate at the source position using \texttt{sosta}, following a procedure similar to that adopted for EP/FXT (Section\,\ref{sec:ep}). When consecutive observations did not yield a detection, we additionally stacked the data to derive more stringent upper limits.

\subsection{Ultraviolet Observations}
\label{sec:uv}
\src\ was observed with the \swift\ UltraViolet/Optical Telescope (UVOT; \citealt{Roming2005}) in image mode using the $uvm2$ (central wavelength 2246\,\AA), $uvw1$ (2600\,\AA), $uvw2$ (1928\,\AA), and $u$ (3465\,\AA) filters. For each observation, the individual exposures were co-added to produce a single summed image per filter. Aperture photometry was then performed with \texttt{uvotsource}, following standard UVOT guidelines\footnote{\url{https://www.swift.ac.uk/analysis/uvot/index.php}}. Source counts were extracted from a circular region of radius 5\arcsec\ centered on the source position, and the background was estimated from a nearby source-free circular region of radius 10\arcsec. When the source was not detected above a signal-to-noise ratio of 3, we report the 3$\sigma$ upper limit on the magnitude returned by \texttt{uvotsource}. The final dataset contains both detections and upper limits, primarily in the $uvm2$ band (Table\,\ref{tab:obsUV}).

\subsection{Optical Observations}
\label{sec:optical}
We used optical imaging and spectroscopic data from ATLAS, SVOM/VT, and SALT. The reduction procedures adopted for these datasets are summarized below.

\subsubsection{ATLAS}
\label{sec:atlas}
We performed forced photometry \citep{Shingles2021} at the source position using the Asteroid Terrestrial-impact Last Alert System (ATLAS) public image archive \citep{Tonry2018}. ATLAS provides photometry in two broadband filters, namely `$c$' (cyan; 4200--6500\,\AA) and `$o$' (orange; 5600--8200\,\AA). We retrieved all available measurements from May 2017 onward, most of which were obtained in the $o$ band.

To retain only high-quality measurements, we excluded detections flagged for known photometric issues, required the centroid coordinates to lie within the inner region of the detector in order to avoid measurements affected by truncated point-spread functions or unreliable local background estimates near the detector boundaries, and imposed cuts on image depth and background quality by requiring both the local 5$\sigma$ limiting magnitude and the sky background to be fainter than 17\,mag.

\subsubsection{SVOM/VT}
\label{sec:svom_opt}
We obtained high-cadence optical photometry of \src\ with the Visible Telescope (VT; \citealt{Qiu2026}) on board SVOM on 2025 March 30--31. A sequence of 100-s exposures was acquired in the $VT_B$ and $VT_R$ bands, corresponding to the blue (4000--6500\,\AA) and red (6500--10000\,\AA) channels of the instrument, respectively, and covering a time span of $\approx$8\,hr. The data were reduced following standard procedures, including bias subtraction, dark correction, and flat-fielding. We then performed differential photometry and applied aperture corrections. Because of the high stellar density in the field, no isolated stars were available to derive reliable aperture corrections directly. To reduce contamination from nearby sources, we adopted a small aperture radius of $r=2.0$\,pix ($\simeq$1.5\arcsec). We applied aperture corrections of $\Delta VT_R \simeq -0.2$\,mag and $\Delta VT_B \simeq -0.4$\,mag, derived from a comparison field with lower stellar density. The relative photometry was calibrated using a bright, isolated comparison star in the same field (R.A. = 17:49:39.62, Decl. = -38:48:42.21; J2000; $VT_B\simeq15.90$\,mag; $VT_R\simeq14.85$\,mag). This procedure may introduce a systematic offset of up to 0.1\,mag in both bands. The full details of the SVOM/VT detections are reported in Table\,\ref{tab:obsVT}.

\subsubsection{SALT}
\label{sec:salt}
Two low-resolution spectra of \src\ were obtained with the Robert Stobie Spectrograph (RSS; \citealt{2003SPIE.4841.1463B,2003SPIE.4841.1634K}) on the Southern African Large Telescope (SALT; \citealt{Buckley2006}). The observations started on 2025 June 1 at 21:56 UTC and 2025 June 5 at 20:14 UTC, with exposure times of 2400\,s in both cases. Observing conditions were clear on both nights, and the guide-star images showed a full width at half maximum (FWHM) between 0.9\arcsec\ and 1.2\arcsec. The PG700 grating provided wavelength coverage from $\sim$3100 to 6770\,\AA\ at a resolving power of $\sim$640.
The CCD pre-processing steps (overscan correction, bias subtraction, gain correction, and cosmic-ray removal) were carried out with the \texttt{pysalt}\footnote{For more details on \texttt{pysalt}, visit \url{http://pysalt.salt.ac.za/}.} package \citep{2010SPIE.7737E..25C}. Wavelength calibration was derived from CuAr arc exposures obtained at the end of each science exposure.

The remaining reduction steps were performed using \texttt{IRAF} and/or \texttt{pyraf}. These included spectral extraction and optimal extraction \citep{1986PASP...98..609H} of the science and standard-star spectra, together with sky subtraction, to derive relative flux distributions.

\subsection{MeerKAT Radio Observations}
\label{sec:meerkat}
We monitored \src\ with the MeerKAT array \citep{Jonas2016} as part of the X-KAT Programme (PI: Fender). The campaign comprised 17 weekly observations obtained between 2025 April 6 and July 27 (Table\,\ref{tab:obsmeerkat}).

All data were acquired in the L band, centered at 1.28\,GHz, with a total processed bandwidth of 0.86~GHz (856--1712~MHz). The correlator provided 4096 frequency channels with an integration time of 8~s per visibility. After initial Radio Frequency Interference (RFI) flagging, the data were averaged spectrally to 1024 channels to facilitate processing. Antenna participation ranged from 62 to 64 (out of 64), yielding baselines up to 7.698~km. Each observation lasted 15~min on source and was bracketed by two 2-minute scans of the secondary calibrator J1726-5529; a single 10-minute scan of the primary calibrator J1939-6342 and a 10-minute scan of the polarisation calibrator J1331+3030 were also obtained.

Semi-automated calibration and imaging were performed with \textsc{OxKAT} \citep{oxkat}, a suite of dedicated \textsc{Python} scripts. Within \textsc{CASA} \citep{casa}, we flagged the outer 100 channels at the band edges, removed autocorrelations and zero-amplitude visibilities, and applied additional time- and frequency-domain RFI flagging. Solutions for the flux-density scale, bandpass, and instrumental delay were derived from the primary calibrator, while residual delays and complex gains were obtained from the secondary calibrator. To model the phase calibrator, we temporarily binned the band into eight equal spectral windows referenced to the flux scale of the primary calibrator. These solutions were then applied to the target field. The target data were subsequently split into a separate measurement set, averaged by a factor of 8 in frequency, and further cleaned of residual RFI using \textsc{Tricolour} \citep{tricolour}.

Imaging was carried out with \textsc{WSClean} \citep{wsclean}. We first generated a wide-field image (field of view $\simeq 1.5~\mathrm{deg}^2$) to construct a deconvolution mask, adopting a Briggs robust parameter of 0.0. This mask was then used to re-image the dataset, followed by direction-independent self-calibration with \textsc{CubiCal} \citep{cubical} to solve for phase and delay terms on 32~s intervals.

No significant emission was detected at the nominal position of \src\ in any individual epoch. We therefore performed a stacking analysis, grouping epochs according to the X-ray-inferred source activity. Specifically, we stacked epochs 1--2, 4--8, 11--13, and 14--17, and we also combined all epochs to produce a deep field image with a total on-source integration time of 4\,h\,15\,min. Even in the deep image, which reaches an average rms noise of 27.9\,$\mu$Jy at the position of \src, no significant detection was obtained. Table\,\ref{tab:obsmeerkat} reports the flux-density upper limits derived from the individual and stacked images.

\section{Data Analysis and Results}
\label{sec:analysis}

\subsection{Localization, Counterpart Searches, and Distance}
\label{sec:position}
\begin{figure}[htb]
\centering
  \includegraphics[width=0.47\textwidth]{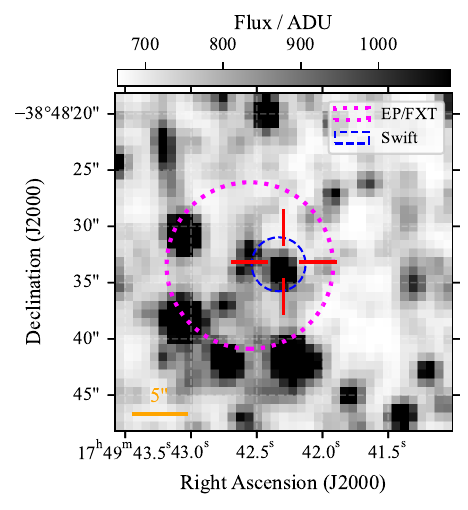}
\vspace{-0.5cm}
 \caption{Zoomed-in ($30\arcsec \times 30\arcsec$) $VT_{\rm R}$-band image extracted from SVOM/VT observations (2025 March 30--31), centered on the optical counterpart of \src. The dotted magenta circle (radius $7.4\arcsec$) and the dashed blue circle (radius $2.4\arcsec$) mark the 90\% confidence X-ray localization uncertainties from EP/FXT and \swift, respectively. The red crosshair indicates the position of the optical counterpart as reported in \gaia\ DR3 (see Section\,\ref{sec:position} for details).}
\label{fig:vt_zoom}
\end{figure}

Figure\,\ref{fig:vt_zoom} shows the optical field around the EP/FXT and \swift\ positions, extracted from SVOM/VT data. The first pointed X-ray observation of the field was carried out using EP/FXT, starting on 2025 March 24 at 06:02:34 UTC -- over 19 days after the initial EP/WXT detection.
An uncataloged X-ray source was detected at a position consistent with the EP/WXT localization: R.A. = 17$^\mathrm{h}$49$^\mathrm{m}$42$\fs$55, Decl. = --38$^{\circ}$48$^{\prime}$33$\farcs$5 (J2000.0), with a positional uncertainty of 7.4$\arcsec$ radius at 90\% confidence level (c.l.).
\swift\ follow-up observations yielded a more precise position via astrometric calibration using field stars in the UVOT images \citep{Evans2009}: R.A. = 17$^\mathrm{h}$49$^\mathrm{m}$42$\fs$33, Decl. = --38$^{\circ}$48$^{\prime}$33$\farcs$4 (J2000.0), with an error radius of 2.4$\arcsec$ (90\% c.l.). The corresponding Galactic coordinates are $l \simeq 352\fdg5$, $b \simeq -5\fdg6$, placing the source close to the direction of the Galactic Center, slightly below the Galactic plane.

Within the \swift\ error circle, two sources are classified as stars in the \gaia\ Third Data Release (DR3; \citealt{GaiaDR32023}). Using the likelihood ratio formalism by \cite{Sutherland1992}, we identify \gaia\ DR3 4036709435120357888 as the most probable optical counterpart of \src\ (for details, see Appendix\,\ref{app:lr_method}). This source is located at R.A. = 17$^\mathrm{h}$49$^\mathrm{m}$42$\fs$29$\pm$0$\fs$00003 and Decl. = --38$^{\circ}$48$^{\prime}$33$\farcs$2$\pm$0$\farcs$0003, with a $G$-band (3300--10500\,\AA) magnitude of $\simeq$19.7\,mag and a color index of $\mathrm{BP} - \mathrm{RP} = 1.06$\,mag. Optical photometry at the \gaia\ position shows long-term variability closely tracking the X-ray emission over the activity period of \src\ (see Section\,\ref{sec:lcurve}). The second candidate shows no significant variability, supporting the identification of \gaia\ DR3 4036709435120357888 as the optical counterpart\footnote{The second \gaia\ source has a magnitude of $G\simeq 20.2$\,mag (see Appendix\,\ref{app:lr_method}) and is much fainter than the transient counterpart visible in Fig\,\ref{fig:vt_zoom}. Any photometric contamination arising from the different apertures used in the optical and UV data reduction (see Sections\,\ref{sec:uv} and \ref{sec:optical}) is expected to be negligible.}.

The position of the \gaia\ counterpart matches a NIR source listed in the DR4.2 release of the VISTA Variables in the Via Lactea (VVV) survey \citep{Saito2012,Minniti2023}, which has average magnitudes of $J \simeq 18.5$\,mag and $H \simeq 18.3$\,mag (Vega). No radio counterpart is detected in existing surveys (see Appendix\,\ref{sec:radio_cutouts}).

\begin{figure}[th]
\begin{center}
\includegraphics[width=0.45\textwidth]{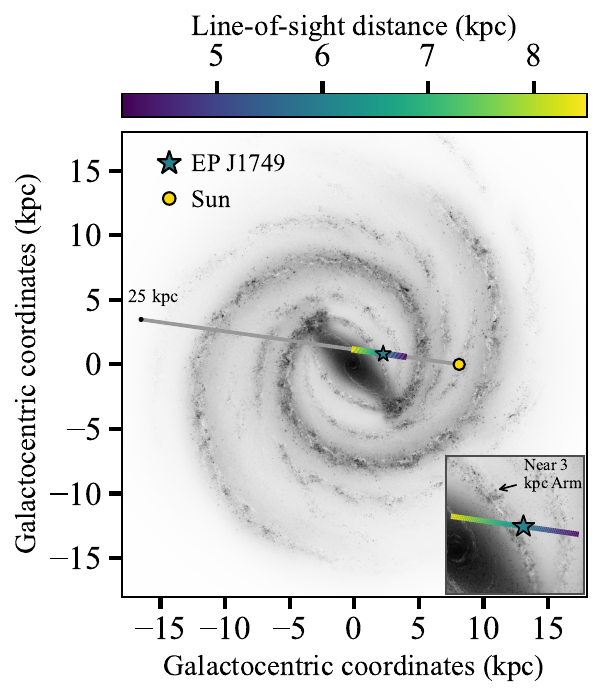}
\vspace{-0.3cm}
\caption{Face-on view of the Galactic disk showing the LoS toward \src\ (illustration credit: NASA/JPL-Caltech/R.~Hurt (SSC/Caltech)). The colored segment traces the LoS over the distance range inferred from the \gaia\ parallax measurement (see text). The position of \src\ is marked with a star symbol. The inset shows a zoomed-in view around
the position of \src.}
\label{fig:ep1749_mw_spirals}
\end{center}
\end{figure}

The parallax of the \gaia\ source is $\varpi = 0.64 \pm 0.51$\,mas, which, based on the probabilistic estimate from the \gaia\ Early DR3 distance catalog \citep{BailerJones2021}, corresponds to a geometric distance of $d = 6.0^{+2.5}_{-1.9}$\,kpc (90\% c.l.). This places \src\ at a location consistent with the Near 3 kpc Arm (see Fig.\,\ref{fig:ep1749_mw_spirals}), at a height of $|z| = 0.61^{+0.25}_{-0.19}$\,kpc above the Galactic plane.
Using the 3-D extinction and column density tool by \citet{Doroshenko2024}\footnote{\url{http://astro.uni-tuebingen.de/nh3d/nhtool}.}, we estimated a line-of-sight (LoS) hydrogen column density $N_{\rm{H}} = (3.0^{+3.0}_{-2.7}) \times 10^{21}$\,cm$^{-2}$ at the \gaia\ position and the assumed distance. This is consistent within the uncertainties with that derived from the modeling of the broadband X-ray spectrum (see Section\,\ref{sec:xrayspec_broadband}). The estimated extinction along the LoS is $E(B-V) = 0.27 \pm 0.04$\,mag. We obtained a consistent value using the recent 3-D dust map of the southern Galactic plane by \citet{Zucker2025}.

For all subsequent timing analyses, the arrival times of photons collected by the various instruments were corrected to the Solar System barycenter using the \gaia-based source position and the JPL planetary ephemeris DE440 \citep{Park2021}. We assumed a source distance of 6\,kpc.

\begin{figure*}[!htb]
\begin{center}
\includegraphics[width=1.0\textwidth]{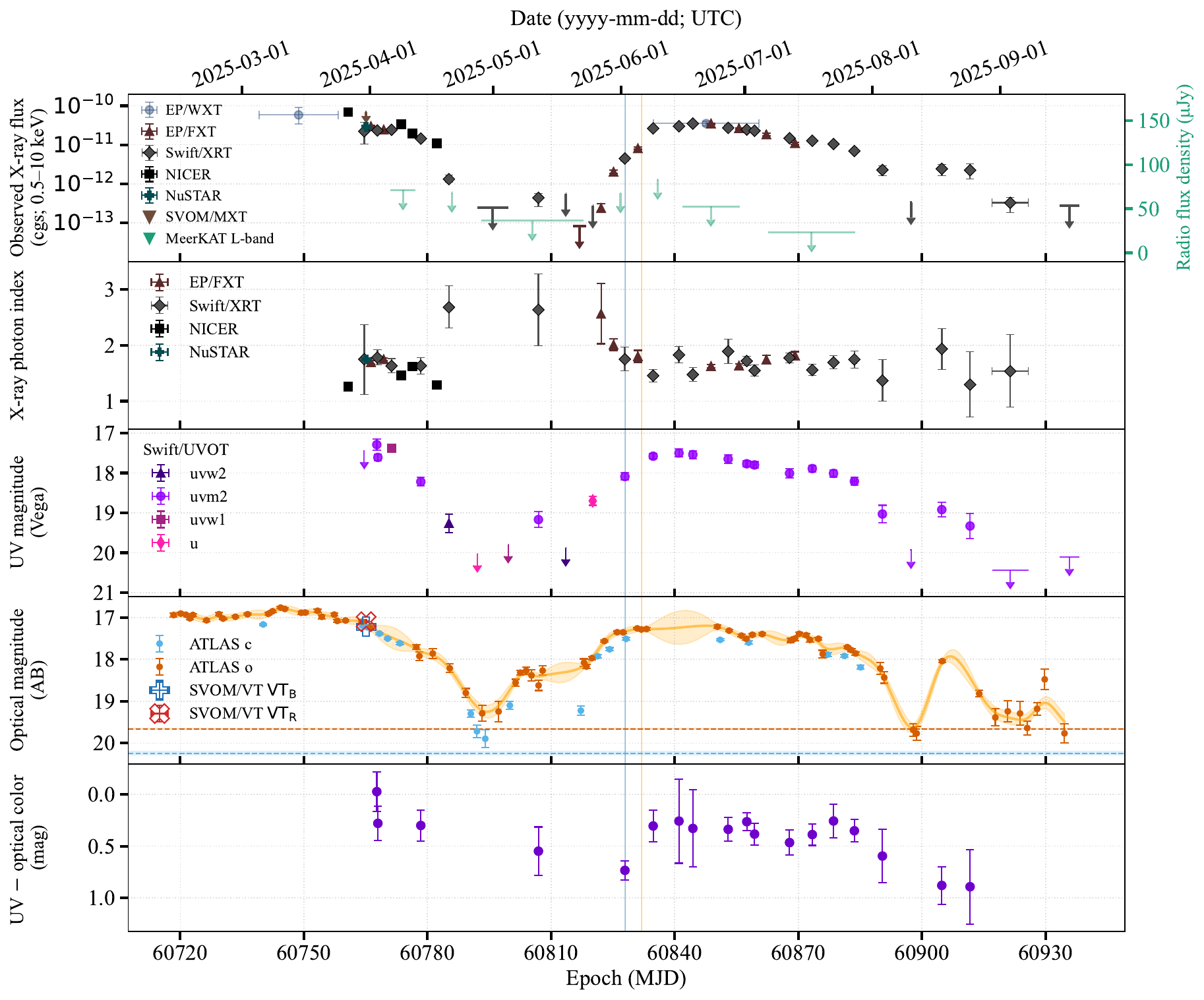}
\vspace{-0.5cm}
\caption{Multi-wavelength temporal evolution of \src.
\emph{Panel 1:} Observed 0.5--10\,keV fluxes in cgs units (\flux) along with the 3$\sigma$ upper limits on the radio flux density in the L-band.
\emph{Panel 2:} X-ray photon index.
\emph{Panel 3:} Multi-filter UV photometry.
\emph{Panel 4:} Optical photometry.
The orange curve and shaded area show the GP interpolation and uncertainty envelope for ATLAS $o$ band data.
Horizontal dashed lines mark the quiescent levels in the ATLAS bands.
\emph{Panel 5:} UV $-$ optical color evolution based on \swift/UVOT $uvm2$ and ATLAS $o$-band data.
Vertical lines indicate the epochs of the SALT spectroscopic observations.}
\label{fig:mw_lc}
\end{center}
\end{figure*}

\subsection{Multi-wavelength Temporal Evolution}
\label{sec:lcurve}
Figure\,\ref{fig:mw_lc} shows the multi-wavelength temporal evolution of \src\ over $\approx$220\,days (from mid-February to mid-September 2025), as observed in the X-ray, UV, and optical bands.
The X-ray flux and photon index are obtained from the spectral fits described in Section\,\ref{sec:xrayspec} and are used here to follow the evolution across bands.
The initial 0.5--10\,keV observed flux was $\approx 5\times10^{-11}$\,\flux\ and remained at comparable levels for about a month following the first X-ray detection on March 5 (MJD 60739). This was followed by a sharp decline of more than two orders of magnitude over $\simeq$20 days, after which the source re-brightened over $\simeq$10--15 days to flux levels comparable to the initial outburst. The X-ray emission subsequently began a second decay, eventually dropping below the \swift/XRT detection threshold by August 10 (MJD 60897). A rebrightening, lasting about two weeks, occurred thereafter. Following this episode, the source transitioned back into quiescence, where it remains at the time of writing.

The first available UV observation, taken on March 30 (MJD 60744) during the initial phase of enhanced X-ray emission, detected the counterpart at a magnitude of $\approx$17.5\,mag (Vega). This detection was followed by a fading trend interspersed with periods of non-detection. Subsequently, a renewed UV brightening was measured along with the onset of the second X-ray outburst, after which the UV emission again began to fade.
Importantly, optical observations revealed that the counterpart was already brightened by February 12 (MJD 60718; the first available measurement after a $\sim$4-month gap in ATLAS observations), several weeks before X-ray activity was first detected. Although the outburst onset epoch remains uncertain, this early optical brightening indicates that the source was already active at the time of the initial EP/WXT detection. The optical data further reveal a dimming by $\sim$2\,mag at the end of the first X-ray outburst, with a subsequent re-brightening accompanying the second outburst. Overall, the optical evolution closely mirrors the behavior seen in both the X-ray and UV bands.

A UV $-$ optical color index was derived using $uvm2$ and $o$-band detections. When no simultaneous measurement was available (offsets $\geq$0.5\,d), we interpolated the $o$-band magnitudes using Gaussian Process (GP) modeling (for a review, see \citealt{Aigrain2023}). Specifically, we modeled the ATLAS $o$-band light curve with a GP employing a Matérn-3/2 kernel combined with constant and white-noise terms. We then used the GP's predictive mean and 1$\sigma$ uncertainty to estimate the optical magnitude at the UVOT epochs. The resulting UV $-$ optical color curve exhibits a blueward shift during the outburst, consistent with the emergence of a hot, blue-emitting component.

No radio emission was detected in any individual observation. To improve sensitivity, we performed a stacking analysis by grouping observations based on the X-ray flux evolution, but still obtained non-detections. The resulting 3$\sigma$ upper limits on the radio flux density range from 20 to 80\,$\mu$Jy at 1.28\,GHz (see top panel of Fig.\,\ref{fig:mw_lc} and Table\,\ref{tab:obsmeerkat}).

\subsection{X-ray Aperiodic Variability and Spectral Trends}
\subsubsection{Power Spectrum Analysis}
\label{sec:PSD_analysis}
To investigate the aperiodic variability of \src, we focused on the \nustar\ observation (2025 March 30--31), which provides both the highest photon statistics and the longest exposure of our campaign. We selected photons in the 3--40\,keV energy range and extracted a time series with 1\,ms resolution. We then computed an averaged power density spectrum (PDS) by dividing the time series into 512-s segments, extracting the power spectrum for each segment, correcting for windowing effects caused by gaps in the data (see \citealt{ElByad2025}), and applying fractional root-mean-square (rms) normalization \citep{Belloni1990}.
We adopted 512-s segments as a compromise between low-frequency coverage and statistical robustness. This choice sets a minimum sampled frequency of $\sim2\times10^{-3}$\,Hz, allowing us to probe the low-frequency regime where broad-band noise and quasi-periodic features are typically observed in LMXBs, while retaining a sufficient number of independent segments for averaging. Repeating the analysis with different segment lengths yielded consistent PDS shapes. The resulting PDS was logarithmically rebinned with a frequency resolution of $\Delta\nu/\nu = 2$\%.

\begin{figure}[!h]
\begin{center}
\includegraphics[width=0.47\textwidth]{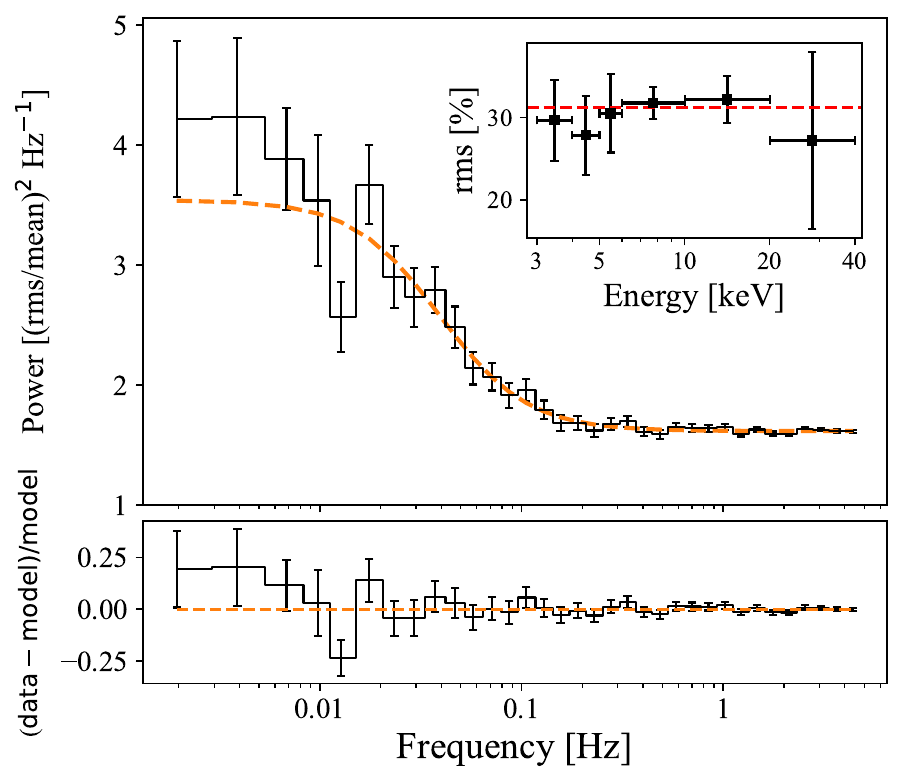}
\vspace{-0.5cm}
\caption{\emph{Top}: logarithmically rebinned average PDS of the \nustar\ 3--40\,keV data (black), fitted with a zero-centered Lorentzian plus constant model (orange dashed curve). \emph{Bottom}: post-fit residuals. The inset in the top panel shows the fractional rms amplitude in the 0.001--0.1\,Hz frequency band as a function of energy, computed over six energy intervals; the horizontal red dashed line marks the mean value.}
\label{fig:nustar_psd}
\end{center}
\end{figure}

Figure\,\ref{fig:nustar_psd} shows the PDS restricted to frequencies $\leq$5\,Hz. At frequencies $\lesssim$0.1\,Hz, the PDS exhibits increasing variability power toward lower frequencies. To characterize this broadband noise, we modeled the PDS using a zero-centered Lorentzian plus a constant white-noise component. We obtained maximum-likelihood estimates by minimizing the chi-squared statistic and assessed parameter uncertainties using a Markov Chain Monte Carlo (MCMC) sampler\footnote{We used the tools included in the \texttt{nDspec} package \citep{Lucchini2025}.}.
We obtained $\chi^2 = 29.37$ for 31 degrees of freedom (dof), and the following best-fit parameters: zero-frequency normalization $N = 1.9 \pm 0.2$ (rms/mean)$^2$\,Hz$^{-1}$, half-width at half-maximum of the Lorentzian $\nu_b = 0.040 \pm 0.004$\,Hz, and a Poisson noise level $C = 1.616 \pm 0.005$ (rms/mean)$^2$\,Hz$^{-1}$. All quoted uncertainties refer to 68\% credible intervals.

After subtracting the Poisson noise level, we measure a fractional rms amplitude of 31$\pm$1\% over the 1--100\,mHz frequency range (where red noise is prominent) in the 3--40\,keV energy band. No significant energy dependence of the rms is detected (see the inset of Fig.\,\ref{fig:nustar_psd}).

\subsubsection{Variability Across Intensity States}
\label{sec:var_flux}
To investigate variability as a function of time and source intensity, we focused on the EP/FXT data (0.5--10\,keV energy range), as they offer the best count statistics across a broad range of epochs (from 2025 April 1 to 2025 July 13) and source intensities (see Table\,\ref{tab:obsX}). From these data, we also computed hardness ratios (HR), defined as $\mathrm{HR} = (H - S)/(H + S)$, where $H$ and $S$ are the net count rates in the 2--10\,keV and 0.5--2\,keV energy bands, respectively.

\begin{figure}[h]
\begin{center}
\includegraphics[width=0.47\textwidth]{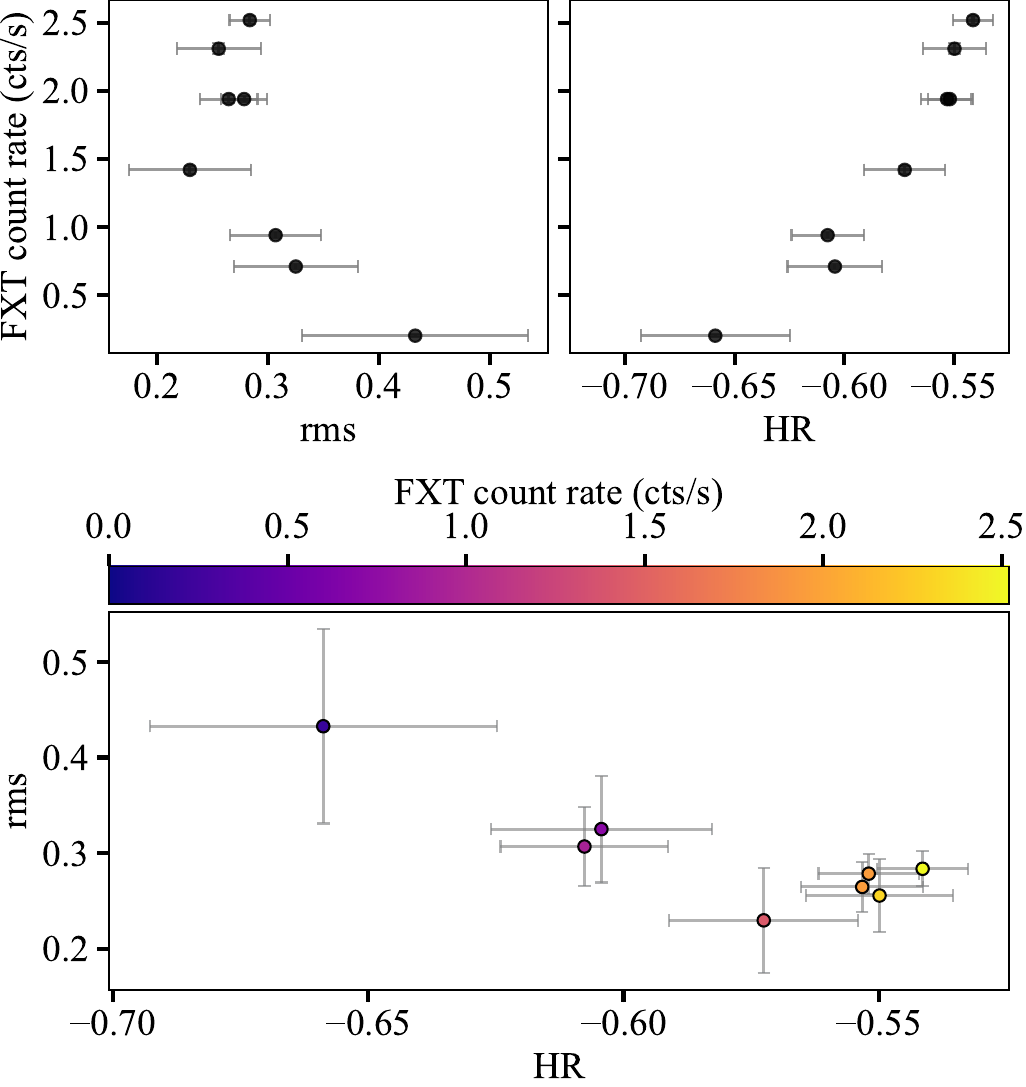}
\vspace{-0.3cm}
\caption{
\emph{Top}: EP/FXT count rate as a function of fractional rms amplitude (\emph{left}) and HR (\emph{right}). \emph{Bottom}: HR versus rms amplitude, with points color-coded by the EP/FXT count rate. Net count rates are in the 0.5--10\,keV energy range, rms amplitudes are in the 1--100\,mHz frequency range. Error bars are shown in gray (in some cases they are smaller than the symbol size).}
\label{fig:hr_rms_cr}
\end{center}
\end{figure}

Figure\,\ref{fig:hr_rms_cr} illustrates how variability and spectral hardness evolve with the source intensity. We quantified these relationships using Spearman's rank correlation test. We found a moderate anti-correlation between rms amplitude and count rate (Spearman coefficient $\rho = -0.61$, $p = 0.11$), although this is not statistically significant at the 5\% level. In contrast, the HR strongly correlates with the count rate ($\rho = +0.97$, $p = 6.5 \times 10^{-5}$). As a consequence, HR and rms amplitude also show a moderate anti-correlation ($\rho = -0.57$, $p = 0.14$), albeit with limited statistical significance.
The results hint at a potential softening of the spectrum when the source becomes fainter.

\subsection{Search for X-ray Pulsations and Bursts}
We searched for coherent X-ray pulsations in the \nicer\ (2025 March 26 -- 2025 April 17) and \nustar\ (2025 March 30--31) datasets, using techniques designed to account for red-noise-dominated power spectra (for details, see Appendix\,\ref{sec:X_searches}). No significant periodic signals were detected, with the most stringent 3$\sigma$ upper limit on the pulsed fraction being $<$6\% in the 1--1000\,Hz range from \nustar. We also performed an acceleration search on \nustar\ data to account for possible Doppler shifts from orbital motion, yielding no detections with the most stringent pulsed fraction limit of $<$26\% (99\% c.l.). Finally, we conducted a burst search across the \nicer, \nustar, and EP/FXT observations on multiple timescales, but found no statistically significant events above a $5\sigma$ threshold (see Appendix\,\ref{sec:X_searches}).

\subsection{X-ray Spectral Analysis}
\label{sec:xrayspec}

\subsubsection{Broadband Spectrum}
\label{sec:xrayspec_broadband}
To investigate the broadband X-ray spectral properties of \src\ during outburst, we performed a joint fit of the EP (FXT-A and FXT-B) spectra acquired on 2025 April 1 and the \nustar\ (FPMA and FPMB) spectra acquired on 2025 March 30--31, during the first outburst. We excluded the SVOM/MXT spectrum from this analysis, as the source was only marginally detected and the data are likely affected by non--X-ray background contamination (see Section\,\ref{sec:svom_x}).
Consequently, the MXT data did not provide additional constraints on the spectral shape beyond those provided by the FXT and \nustar\ observations.

We jointly fit the spectra using the \texttt{XSPEC} spectral fitting package \citep{Arnaud1996}. Interstellar absorption was modeled with the Tübingen-Boulder model (\texttt{TBabs}; \citealt{Wilms2000}), using photoelectric cross-sections from \citet{Verner1996}. We used the Comptonization model \texttt{nthComp} \citep{Zdziarski1996, Zycki1999} to describe the broadband spectrum and assumed that seed photons originate from a cool, disk-like blackbody component. We also included a multiplicative constant (\texttt{const}) to account for both the cross-calibration offsets between instruments and the potential flux variations due to the non-strict simultaneity of the EP and \nustar\ observations. The constant was fixed to unity for the FXT-A data and allowed to vary for the three other instruments. The model provided an excellent fit to the data (reduced $\chi^2$ of $\chi^2_r$ = 0.95 for 546 dof; null hypothesis probability of 78\%), with residuals showing no systematic trends across the 0.5--40\,keV energy range.

\begin{figure}[!ht]
\begin{center}
\includegraphics[width=0.47\textwidth]{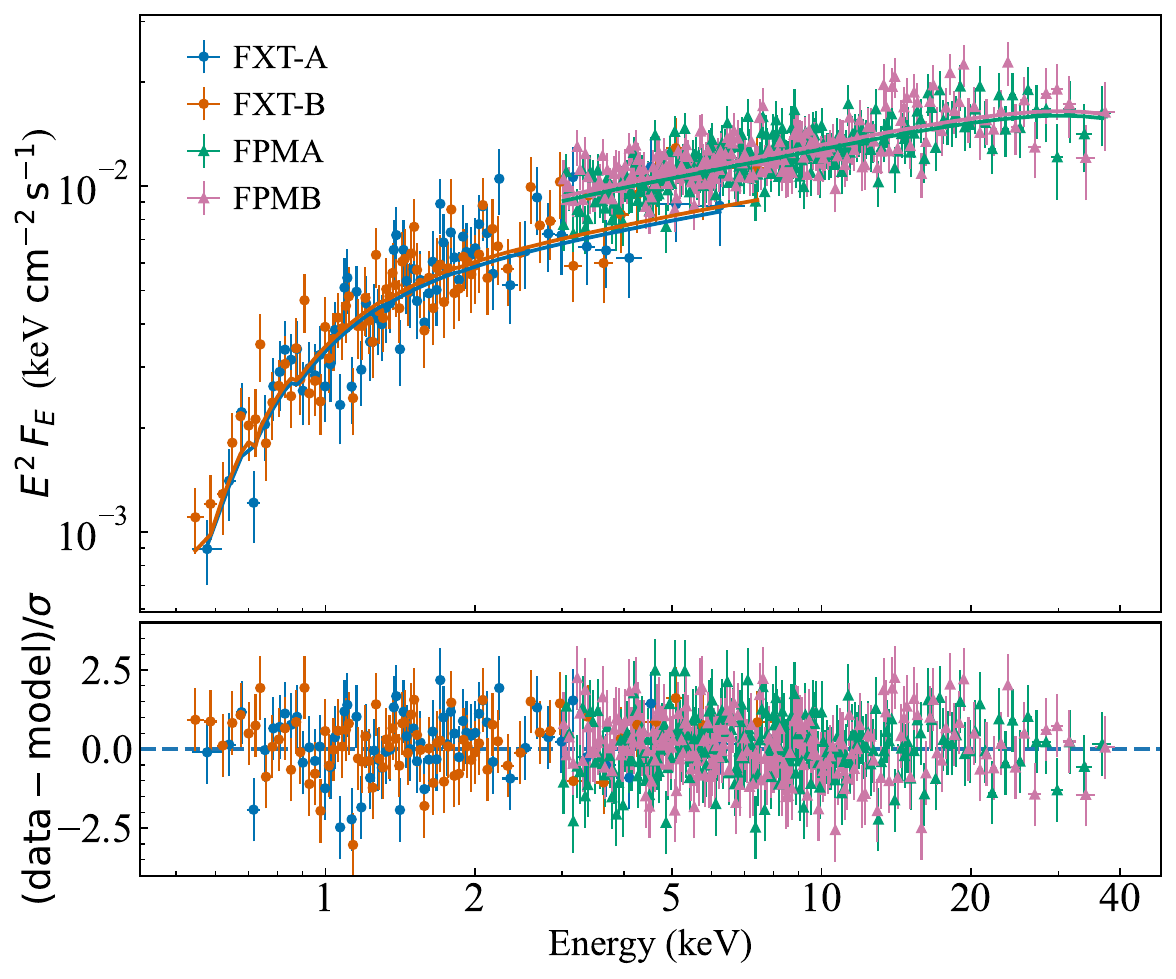}
\vspace{-0.5cm}
\caption{\emph{Top}: unfolded X-ray spectra of \src\ extracted over the 0.5--40\,keV energy range from quasi-simultaneous EP/FXT and \nustar\ observations, along with the best-fit absorbed thermal Comptonization model (solid curves; for details, see Section\,\ref{sec:xrayspec_broadband}). \emph{Bottom}: post-fit residuals. Data from each instrument are shown in different colors. The spectra have been rebinned for visual purposes.
}
\label{fig:broadband_spectrum}
\end{center}
\end{figure}

Despite the lack of evidence for an additional soft thermal component and/or reflection features in the residuals, such as an Fe\,K$\alpha$ emission line (see Fig.\,\ref{fig:broadband_spectrum}), we investigated the presence of additional spectral components. We first included a multicolor disk blackbody (\texttt{diskbb}) to the model, but both its temperature and normalization were completely unconstrained by the fit, indicating that this component was not statistically required. We also attempted to replace \texttt{nthComp} with the self-consistent reflection model \texttt{relxillCp} \citep{Garcia2014}, which accounts for both the Comptonized continuum and its reflection off the disk. While the model describes well the Comptonization spectrum, all parameters related to the reflection component, e.g., the reflection fraction, disk ionization, and inner disk radius, remained completely unconstrained. We therefore adopted \texttt{nthComp} alone to describe the continuum.

Table\,\ref{tab:broadband} lists the results of the spectral fit. We obtained the following best-fit parameters: $N_{\rm{H}} = (2.7\pm0.2) \times 10^{21}$\,cm$^{-2}$, fully consistent within uncertainties with the expected Galactic value along the LoS (see Section\,\ref{sec:position}); a photon index of the Comptonizing component of $\Gamma=1.76\pm0.01$; and a temperature for the electron population of $kT_e=17^{+11}_{-3}$\,keV. We derived a 3$\sigma$ upper limit on the seed photon temperature of $<$0.1\,keV.
These results show that the broadband emission is dominated by thermal Comptonization of very soft seed photons. The low inferred seed-photon temperature and the absence of a directly detected thermal component are consistent with a cool disk whose emission lies largely below the observed X-ray band.

\begin{deluxetable}{lc}[!h]
\tablewidth{0pt}
\tablecaption{Results of the fits to the broadband (0.5--40\,keV) X-ray spectrum of \src\ using an absorbed thermal Comptonization model (\texttt{const*TBabs*nthComp}).\label{tab:broadband}}

\tablehead{
  \colhead{Parameter} &
  \colhead{Value}
}
\startdata
$C_{\rm FXTB}$                             & 1.03$\pm$0.03    \\
$C_{\rm FPMA}$                             & 1.33$\pm$0.05   \\
$C_{\rm FPMB}$                             & 1.37$\pm$0.05   \\
\nh\ (10$^{21}$\,cm$^{-2}$) 	           & 2.7$\pm$0.2  	 \\
$\Gamma$ 	                               & 1.76$\pm$0.01   \\
$kT_e$ (keV) 	                           & 17$^{+11}_{-3}$  \\
$kT_{\rm disk}$ (keV)                      & $<$0.1 		 \\
Observed Flux   ($\times10^{-11}$\,\flux)  & 5.2$\pm$0.2  \\
Unabsorbed Flux ($\times10^{-11}$\,\flux)  & 5.7$\pm$0.2  \\
Luminosity ($\times10^{35}$\,\lum)         & 2.5$\pm$0.1  \\
\hline
$\chi^2_r$ (dof)                           & 0.95 (546)
\enddata
\tablecomments{Uncertainties are at the 1$\sigma$ c.l.; the upper limit is at 3$\sigma$. The fluxes and the luminosity refer to the 0.5--40\,keV energy range. The luminosity was computed assuming a distance of 6\,kpc.}
\end{deluxetable}

\subsubsection{Spectral Evolution}
\label{sec:spec_evol}
To investigate the evolution of the spectral properties over the course of our observing campaign, we fitted all spectra with an absorbed power-law model, fixing $N_\mathrm{H}$ to the Galactic value along the LoS at 6\,kpc \citep{Doroshenko2024}, consistent with that obtained from the broadband fit. Spectral fitting was performed using Cash statistics \citep{Cash1979,Kaastra2017}. Observed and absorption-corrected fluxes in the 0.5--10\,keV band were measured using the \texttt{cflux} model. For epochs corresponding to non-detections, we estimated flux upper limits using \texttt{WebPIMMS}\footnote{\url{https://heasarc.gsfc.nasa.gov/cgi-bin/Tools/w3pimms/w3pimms.pl}.} assuming an absorbed power-law model with $\Gamma$ = 2.5, consistent with spectral fitting results at similar flux levels. The results are reported in Table\,\ref{tab:obsX} and shown in the two upper panels of Fig.\,\ref{fig:mw_lc}.
During the initial outburst phase, the X-ray photon index remained in the range $\simeq$1--2, then mildly softened as the flux declined, and subsequently hardened again during a second outburst episode. This behavior is consistent with the results of the hardness-ratio analysis (see Section \ref{sec:var_flux}). However, the limited photon statistics in the faintest intervals prevent us from determining whether the observed softening reflects the emergence of a thermal component in the spectrum.

\subsection{Short-term Optical Variability}
The optical counterpart was detected in all SVOM/VT exposures collected on 2025 March 30--31 during the first outburst. The average magnitudes were VT$_{\rm B}$ = 17.22 $\pm$ 0.02\,mag and VT$_{\rm R}$ = 17.10 $\pm$ 0.02\,mag (AB). On the native 100-s cadence of the VT light curves, the VT$_{\rm B}$ data show no significant short-timescale variability, with a 3$\sigma$ upper limit on the fractional rms variability amplitude of $F_{\rm var}<1.5$\%, whereas the VT$_{\rm R}$ light curve shows weak variability with $F_{\rm var}= 2.2 \pm 0.1$\%. 
Lomb–Scargle periodograms, computed over timescales accessible to the VT sampling (from $\sim$0.05\,hr to $\sim$8\,hr), reveal no periodic modulation with semi-amplitudes $\geq 0.025$\,mag at the 99\% c.l. The VT$_{\rm B}$ -- VT$_{\rm R}$ color also remained roughly constant, indicating minimal color evolution.

The SVOM/VT observations were simultaneous with the \nustar\ observations. 
However, no significant optical/X-ray correlation was detected (for details, see Appendix\,\ref{sec:zdcf}). We note that the $\sim$100\,s VT sampling limits sensitivity to rapid correlated variability.

\subsection{Optical Spectroscopy}
\begin{figure*}[!th]
\begin{center}
\includegraphics[width=1.0\textwidth]{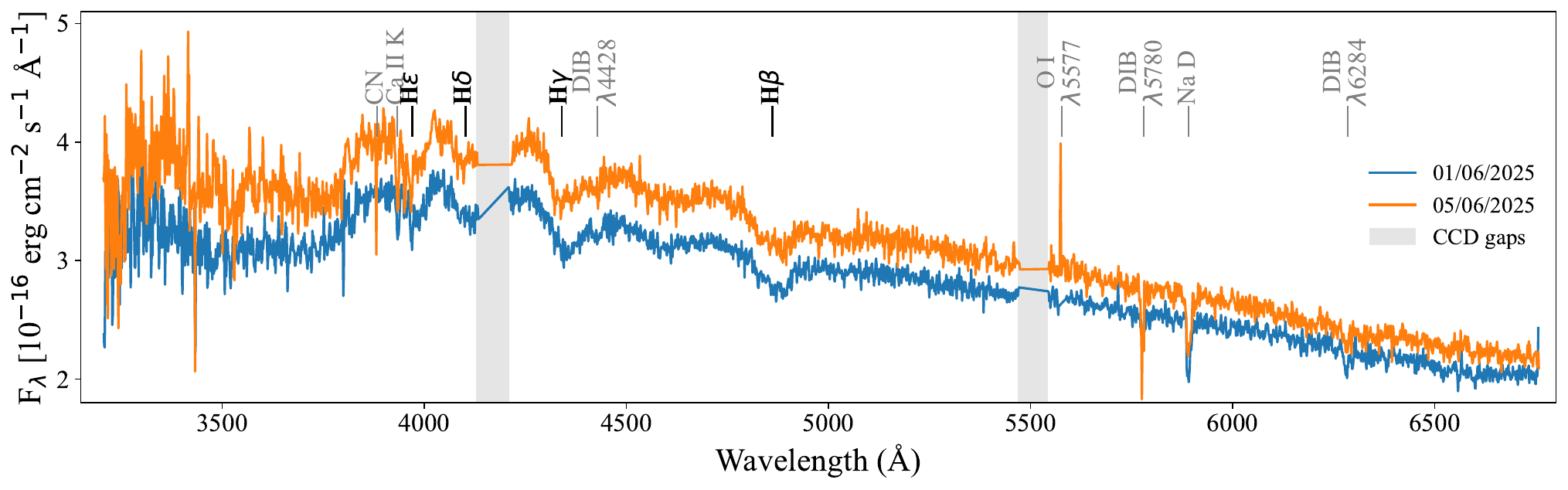}
\vspace{-0.6cm}
\caption{Flux-calibrated spectra of the optical counterpart to \src\ obtained with SALT/RSS on 2025 June 1 (cyan) and June 5 (orange). The spectra show a blue continuum with multiple absorption features, including Balmer lines (H$\epsilon$, H$\delta$, H$\gamma$, H$\beta$; black labels), as well as interstellar absorption from the CN molecular band, the \ion{Ca}{2}~H\&K doublet, the Na\,D doublet, and the diffuse interstellar bands (DIBs) at 4428, 5780, and 6284\,\AA\ (gray labels). A narrow [O\,I] sky emission line at 5577\,\AA\ is visible in the June 5 spectrum. The gray shaded regions mark the wavelength gaps corresponding to the separations between the three CCDs of the RSS mosaic.}
\label{fig:salt_spec}
\end{center}
\end{figure*}

Figure~\ref{fig:salt_spec} shows the optical spectra of \src, obtained on 2025 June 1 and 5, covering the 3300--6800\,\AA\ wavelength range. Both spectra display a blue continuum with prominent Balmer absorption features (H$\beta$, H$\gamma$, H$\delta$, H$\epsilon$), while no significant feature is detected at the wavelength of H$\alpha$. A broad depression is observed in the bluest portion of the spectra, consistent with the Balmer jump at 3646\,\AA\ and with the blending of higher-order Balmer absorption lines.

To measure the line properties, we normalized each spectrum using a local continuum fit around each line, modeled with low-order polynomials and excluding the line core. Equivalent widths (EWs) and FWHMs were then measured from the normalized line profiles. Uncertainties were estimated via MCMC simulations, in which the flux was perturbed according to the per-pixel uncertainties and the continuum refitted in each realization. The standard deviation of the resulting EW and FWHM distributions was adopted as the uncertainty. The results are summarized in Table\,\ref{tab:lines}. The H$\beta$/H$\gamma$ absorption-EW ratio is $1.0 \pm 0.2$ on 2025 June 1 and $1.1 \pm 0.3$ on 2025 June 5, indicating that the relative strengths of the two lines remain consistent within the uncertainties between the two epochs.

\begin{deluxetable}{cccc}
\tablewidth{0pt}
\tablecaption{Properties of the Balmer absorption features in SALT spectra of the optical counterpart to \src\ during the rising phase of the second outburst.\label{tab:lines}}
\tablehead{
  \colhead{Line} &
  \colhead{Epoch} &
  \colhead{EW (\AA)} &
  \colhead{FWHM (km\,s$^{-1}$)}
}
\startdata
H$\gamma$     & 01~Jun & $6.5 \pm 1.3$ & $4831 \pm 626$ \\
H$\beta$      & 01~Jun & $6.6 \pm 0.8$ & $5487 \pm 530$ \\
H$\alpha$     & 01~Jun & $<$1.6 & -- \\
\hline
H$\gamma$     & 05~Jun & $4.8 \pm 0.8$ & $4983 \pm 472$ \\
H$\beta$      & 05~Jun & $5.3 \pm 1.3$ & $5399 \pm 706$ \\
H$\alpha$     & 05~Jun & $<$1.5 & -- \\
\enddata
\tablecomments{Uncertainties are at $1\sigma$ c.l. Upper limits are at $3\sigma$ c.l.}
\end{deluxetable}

\subsection{Multi-band Constraints on Pre-Outburst Activity}
\label{sec:pre-outburst}
Before March 2025, the field of \src\ was observed in soft X-rays by \ros\ (during its all-sky survey), \xmm\ (in three separate Slew Survey observations), and \emph{eROSITA} (according to the eROSITA-DE Data Release 1 archive; see \citealt{Merloni2024}).
No emission was detected from \src\ in any of these observations. Table\,\ref{tab:xobs_archive} lists these observations and the 3$\sigma$ upper limits on the net count rates at the source position, obtained using the \texttt{HIgh-energy LIght-curve GeneraTor} tool (\texttt{HILIGT}\footnote{\url{https://xmmuls.esac.esa.int/hiligt/}}; \citealt{Saxton2022,Konig2022}). The table also reports the corresponding flux upper limits estimated using \texttt{WebPIMMS}, assuming an absorbed power-law model with $N_{\rm H} = 3\times10^{21}$\,cm$^{-2}$ and $\Gamma = 2.5$. The most constraining upper limit on the unabsorbed flux was derived from the \emph{eROSITA} observation (performed on 2020 March 30), $F_X < 3\times10^{-13}$\,\flux\ (0.5--10\,keV), corresponding to a luminosity of $L_X < 1.3\times10^{33}~d_6^2$\,\lum\ (where $d_6$ is the source distance rescaled at 6\,kpc). However, this is not the most constraining limit on the quiescent X-ray emission from \src, as a deeper limit of $F_X < 1.4\times10^{-13}$\,\flux\ was obtained by stacking the EP/FXT observations collected after the first outburst, on 2025 May 20 and 23 (total exposure: 6.3\,ks; see Table\,\ref{tab:obsX}), yielding $L_X < 6\times10^{32}~d_6^2$\,\lum.

To investigate potential pre-outburst optical activity of \src, we systematically analyzed ATLAS photometric data collected before the first outburst. We searched for both isolated brightening events and sustained episodes of enhanced flux. Apart from a few single-epoch artifacts associated with spurious measurements, no significant brightening lasting longer than $\sim$1\,d with an amplitude $\gtrsim$0.5\,mag was detected. We therefore found no compelling evidence for pre-outburst activity in the ATLAS coverage since 2017 May 6. A Lomb–Scargle analysis of the light curves likewise revealed no periodic variability in the quiescent emission (for details, see Appendix\,\ref{sec:periodsearch}).

\section{Discussion}
\label{sec:discussion}

The 2025 outburst of \src\ unveiled a previously unknown Galactic transient, characterized by marked temporal evolution from X-rays to optical. Our monitoring campaign demonstrates that the source underwent two distinct episodes of enhanced activity over a span of $\approx$7 months, during which the X-ray flux varied by more than three orders of magnitude. This behavior was accompanied by correlated brightening and fading of the UV and optical counterpart. Overall, the observed multi-wavelength properties strongly suggest that the transient was powered by accretion onto a compact object in a LMXB.

In the following, we examine the emission properties to investigate the nature of the accretor and the physical mechanisms governing the outburst, focusing first on the X-ray spectral and timing characteristics, and then analyzing the optical and UV behavior, as well as the radio emission. We finally address the origin of the multi-episodic activity and the classification of the source as a BH VFXT candidate.

\subsection{X-ray Emission Properties}
\label{sec:Xray_properties}

\subsubsection{Outburst Evolution and VFXT Context}
In outburst, the source consistently displayed a hard X-ray spectrum ($\Gamma \simeq 1$--$2$) and a low peak luminosity of $L_{\rm X} \approx 5 \times 10^{35}~d_6^2$\,\lum\ in the 0.5--10\,keV energy band (more than three orders of magnitude above the quiescent level). This behavior makes \src\ a compelling candidate member of the class of VFXTs. As the flux decreased, its X-ray spectrum softened, with $\Gamma$ increasing up to $\approx$2.5 (see Fig.\,\ref{fig:mw_lc}).
From the deepest pre-outburst and post-outburst non-detections (Section\,\ref{sec:pre-outburst}), we infer a 0.5--10\,keV quiescent luminosity limit of $L_{\rm X,q} < 6\times10^{32}~d_6^2$\,\lum. For comparison, quiescent BH LMXBs typically span $L_X \sim 10^{30}$--$10^{33}$\,\lum, while NS systems more commonly occupy $L_X \sim 10^{32}$--$10^{34}$\,\lum\ (e.g., \citealt{Bahramian2023,Kalemci2022,DiSalvo2023} and references therein). Our upper limit therefore lies in the region where the two populations overlap and, by itself, does not uniquely distinguish the nature of the accretor, owing to the substantial overlap and intrinsic scatter between the two populations; however, it is fully consistent with a BH system in quiescence.

With a peak X-ray luminosity of a few $\times10^{35}$\,\lum\ and two outbursts plus a rebrightening spread over $\approx$220\,d, \src\ lies near the bright end of the VFXT regime while exhibiting a long-lived, multi-episodic activity pattern. The absence of any comparable optical activity in the ATLAS archive since 2017 implies a duty cycle $\lesssim$10\% on multi-year timescales. Such low-luminosity, low-duty-cycle behavior strengthens the case for a larger, undercounted population of VFXTs that are difficult to detect in all-sky surveys and are often missed without sensitive triggers such as those provided by EP.

\subsubsection{$\Gamma$--$L_{\rm X}$ Diagnostic: NS vs BH}
We compared the X-ray spectral behavior of \src\ with that of well-studied NS and BH LMXBs, focusing on the correlation between $\Gamma$ and $L_{\rm X}$. We employed a Monte Carlo approach for the statistical comparison (see also \citealt{Stoop2021}). For each data point, we sampled values of $\Gamma$ and $\log_{10}(L_X / \mathrm{erg\,s^{-1}})$ from independent Gaussian distributions centered on their measured values, with standard deviations equal to their 1$\sigma$ uncertainties. To account for sampling variability, we resampled each synthetic dataset with replacement -- i.e., by randomly selecting data points from the set -- until the new dataset matched the original sample size. We then fitted a linear relation, $\Gamma = a\bigl[\log_{10}\bigl(L_X/\mathrm{erg\,s^{-1}}\bigr) - 34\bigr] + b$, to each resampled set. This procedure was repeated $10^5$ times for each population, generating posterior distributions for both the slope ($a$) and the offset ($b$). For \src, we obtained median values of $a=-0.2\pm0.2$ and $b=1.9\pm0.2$ (68\% credible intervals).

\begin{figure*}[!th]
\begin{center}
  \includegraphics[width=1.0\textwidth]{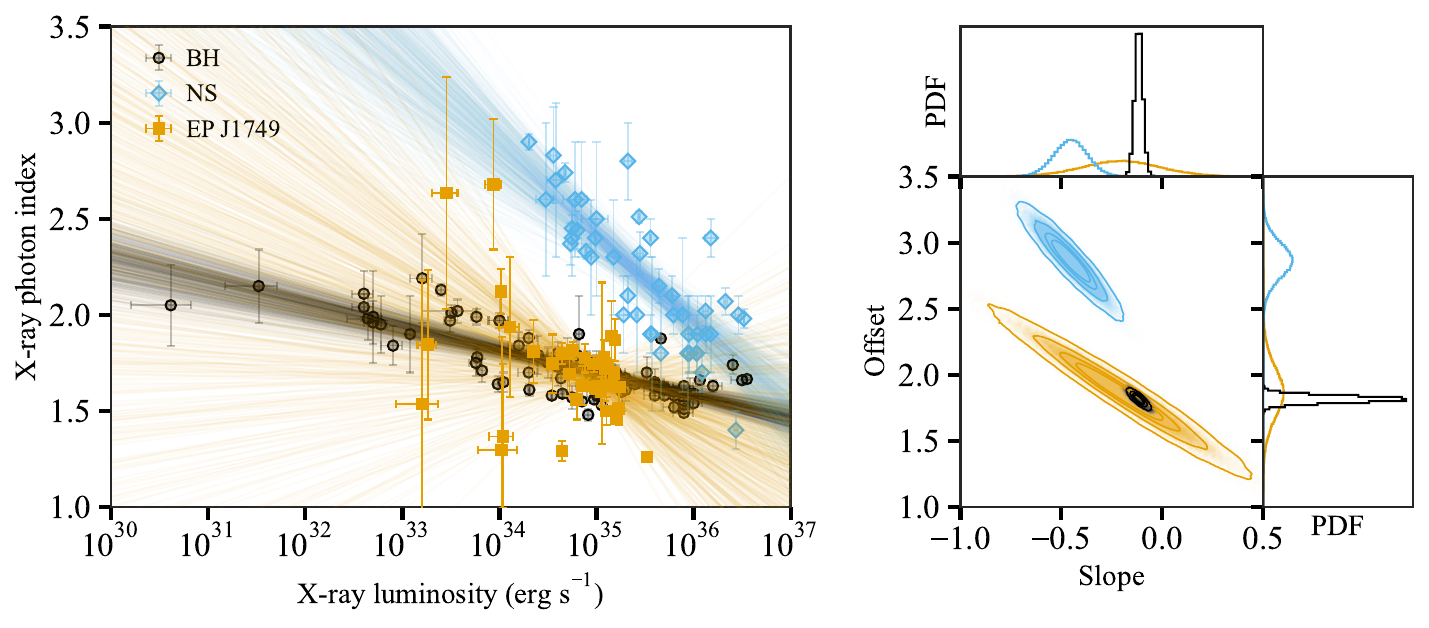}
\vspace{-0.5cm}
 \caption{\emph{Left}: X-ray photon index versus 0.5--10\,keV luminosity, with 1$\sigma$ error bars, for BH systems (black circles), NS systems (cyan diamonds), and \src\ (orange squares). Overlaid are 600 linear regression lines randomly drawn from the Monte Carlo analysis, color-coded by population. The NS and BH data are taken from \citet{Wijnands2015, Plotkin2017, Shaw2021}.
\emph{Right}: Posterior distributions of the regression parameters from $10^5$ Monte Carlo realizations per population. Contours represent kernel density estimates enclosing 1\%, 10\%, 30\%, and 50\% of the maximum density. The top and right sub-panels show the marginalized posterior probability density functions for the slope and offset, respectively.}
\label{fig:gamma_lx}
\end{center}
\end{figure*}

The left panel of Fig.\,\ref{fig:gamma_lx} displays the $\Gamma$--$L_X$ scatter plot together with 600 randomly selected regression lines, color-coded by population. The right panel shows the posterior distributions on the $(a,b)$ plane. \src\ overlaps with the BH posterior distribution in both slope and offset, while it lies significantly outside the region occupied by NSs. We repeated the analysis assuming distances of 4 and 8\,kpc, consistent with the uncertainties in the \gaia\ parallax measurement (see Section\,\ref{sec:position}), and the outcome did not change. To appear more consistent with the NS correlation rather than with the BH track, \src\ would need to be located at a distance $\gtrsim25$\,kpc. This is beyond the expected extent of the Galactic stellar disk along this LoS (see Fig.\,\ref{fig:ep1749_mw_spirals}). We thus conclude that \src\ occupies the same $\Gamma$--$L_X$ parameter space as BH LMXBs, supporting the presence of a BH accretor.

\subsubsection{Constraints on the Accretion Flow}
The analysis of the broadband X-ray spectrum shows that the emission from \src\ is dominated by thermal Comptonization of soft photons. We derive a moderately hard photon index ($\Gamma \simeq 1.8$) and a comparatively low electron temperature for the Comptonizing plasma ($kT_e = 17^{+11}_{-3}$\,keV), implying a spectral rollover within the \nustar\ bandpass. Such low $kT_e$ values have been reported in BH VFXTs (e.g., \citealt{Armas-Padilla2014a}), although BHs in the canonical hard state more commonly exhibit higher electron temperatures of $\sim$30--200\,keV. In contrast, hard-state NS spectra frequently show cooler Comptonizing plasmas, plausibly due to additional cooling from seed photons originating at the stellar surface and boundary layer, with typical electron temperatures in the 15--25\,keV range, although higher values are also observed (e.g., \citealt{Burke2017} and references therein). Given the large uncertainty on our derived $kT_e$ and the lack of spectral coverage above $\sim$40\,keV, we do not regard this measurement as a decisive diagnostic of the nature of the accretor.

The low temperature of the seed photons ($<$0.1\,keV at 3$\sigma$) suggests that the accretion disk is truncated far from the compact object, making any thermal disk component undetectable in the X-ray spectrum and limiting the supply of seed photons for Comptonization. Additional evidence for disk truncation is provided by the broad-band properties of the PDS of the \nustar\ data shown in Fig.\,\ref{fig:nustar_psd}. Broad-band PDSs can usually be well fitted with a number of zero-centered Lorentzians \citep{Done2007}. However, the dominant Poisson noise in the \nustar\ data of \src\ limits the detectable fraction of the intrinsic variability. Here we assume that the zero-centered Lorentzian fit presented in Section\,\ref{sec:PSD_analysis} corresponds to $\nu_{b}$ \citep{Done2007}. According to the propagating accretion-rate fluctuations model \citep{Lyubarskii1997,Arevalo2006}, broad-band variability in accreting systems is produced by fluctuations in the viscosity that originate at large radii and propagate inward through the accretion flow. In this framework, $\nu_{b}$ traces the viscous frequency at the inner edge of the accretion flow, which evolves during state transitions \citep{Done2007} and shows correlations with other broad-band components and quasi-periodic oscillations \citep{Wijnands1999}.
In such a case, the viscous frequency at the truncation radius can be expressed as $\nu_{b} = \alpha \left(H/R \right)^{2} \nu_{\phi}$, where $\alpha$ is the (dimensionless) Shakura--Sunyaev viscosity parameter \citep{Shakura1973}, $H/R$ is the accretion disk scale height and $\nu_{\phi}=(GM/R^{3})^{1/2}$ is the Keplerian angular frequency evaluated at the truncation radius\footnote{Some authors instead adopt the orbital frequency, which would rescale the inferred radius by a factor $(2\pi)^{-2/3}$ at fixed $\nu_b$, $\alpha$, and $H/R$ (see discussion by \citealt{Veresvarska2023}).}.
Assuming a $\sim$10$M_{\odot}$ BH, $\alpha \sim 0.1$ and $H/R \sim 0.1 - 0.3$ (bracketing the cases of a geometrically thin disk and a moderately thick hot flow; \citealt{Done2007}), and adopting $\nu_{b} = 0.04$\,Hz, the inferred truncation radius is in the range 64--275\,$r_{g}$ (960--4130\,km). Here, $r_g = GM/c^2 \simeq15$\,km is the gravitational radius for a $10\,M_\odot$ BH. These values are well within the expected, and observed, range for disk truncation (e.g., \citealt{Done2007,Furst2016,Xu2017,Wang-Ji2018,Zdziarski2022}).

  All sampled points in the rms--intensity diagram from the EP/FXT data (see top-left panel of Fig.\,\ref{fig:hr_rms_cr}) lie in a high-variability regime, with fractional rms amplitudes of $\simeq$20--40\% in the 1--100\,mHz band. Transient BH LMXBs in their hard states typically occupy this high-rms branch, whereas transitions to softer states are accompanied by a marked drop in rms (\citealt{MunozDarias2011}; for reviews on accretion states of BH systems, see, e.g., \citealt{Homan2005,Remillard2006,Belloni2010,Belloni2016,Kalemci2022}). Similar patterns are observed in transient NS LMXBs at sub-Eddington luminosities, although their detailed tracks can differ due to the additional contribution from the stellar surface and boundary layer \citep{MunozDarias2014}. The high rms amplitudes therefore independently support the conclusion that \src\ remained in a faint hard state throughout the outburst.

The progressive spectral softening observed as the source faded is consistent with the behavior reported in several BH LMXBs \citep[e.g.,][]{Corbel2006,Plotkin2013,Homan2013,ArmasPadilla2013,Plotkin2017,Shaw2021}. This phenomenon is often discussed in the context of accretion flows at very low mass accretion rates, where the geometrically thin, optically thick accretion disk is expected to recede from the compact object, while the inner flow becomes hot, geometrically thick, and optically thin. Such a configuration is commonly described in terms of a radiatively inefficient or advection-dominated accretion flow (RIAF/ADAF; see reviews by \citealt{Done2007,Poutanen2014,Yuan2014}; see also \citealt{Esin1997,Gardner2013,Yang2015}).
As the accretion rate declines, the disk cools and its characteristic emission shifts to lower energies (from soft X-rays toward the extreme-UV, UV, or optical bands), reducing the flux of seed photons available for Comptonization. At the same time, the decreasing optical depth of the hot flow reduces the average number of scatterings. Together, these effects reduce the efficiency in the production of high-energy photons, weaken the hard X-ray tail, and result in a progressively softer power-law spectrum.

Taken together, these spectral and timing constraints point to a persistently truncated accretion geometry, in which a cool outer disk coexists with a hot inner Comptonizing flow throughout the outburst. The low seed-photon temperature, the characteristic variability timescale, the high fractional rms, and the gradual softening during the decay are all naturally explained in the framework of accretion at very low rates through a radiatively inefficient hard-state flow. Although none of these diagnostics alone can unambiguously distinguish between a BH and an NS accretor, their combination strengthens the interpretation of \src\ as a BH-like VFXT that never underwent a transition to a soft state.

\subsection{Optical and UV Emission Properties}

\subsubsection{Quiescent Optical Counterpart}
\label{sec:quiescent_optical}

 To assess the nature of the optical counterpart in quiescence, we examined the available archival optical/NIR photometry under the working assumption that the emission is dominated by the companion star. The broad-band \gaia\ DR3 colors of the identified counterpart, together with the $J$- and $H$-band measurements reported in Section\,\ref{sec:position}, are broadly consistent with a mid-G main-sequence-like companion located at the distance and reddening inferred above. As an additional check, we compared the broad-band photometry with a simple reddened blackbody of effective temperature $T_{\rm eff}\simeq 5.7\times10^3\,\mathrm{K}$, adopting $E(B-V)=0.27$ and the Milky Way-average extinction curve of \citet{Gordon2024} with fixed $R_V=3.1$ for the extinction correction (see Section\,\ref{sec:position}). We fixed the normalization of the model by matching the dereddened $H$-band flux, using the reddest available band as a practical anchor for the SED comparison. With this scaling, the model reproduces the optical/NIR SED within $\sim$0.1--0.25\,mag. These results show that the current sparse photometry is broadly compatible with a companion-dominated quiescent SED; they do not rule out a contribution from residual accretion or a jet.

This comparison should nevertheless be regarded as indicative only. The SED is sampled in only a few broad bands, and the blackbody is used as an approximate representation of the companion continuum. Residual accretion, irradiation, and tidal distortion may affect the observed colors and luminosity, and we therefore refrain from deriving detailed stellar parameters such as metallicity or age. Time-resolved optical spectroscopy in quiescence will be essential to detect photospheric absorption features, quantify any veiling, and robustly establish the nature of the companion star.

 \subsubsection{Origin of the Optical/UV Outburst Emission}
\label{sec:corr}
During outburst, the optical emission displays a prominent blue continuum. To investigate its origin, we examined the correlation between optical and X-ray emission.
We first extracted daily averages in the ATLAS $o$-band and corrected them for interstellar extinction by adopting $E(B-V)=0.27$ (Section\,\ref{sec:position}) together with a standard Milky Way extinction curve from \citet{Gordon2024} with fixed $R_V=3.1$ (see also \citealt{Gordon_2009, Fitzpatrick_2019, Gordon_2021, Decleir_2022}).
We focused on the $o$-band because it offers the densest photometric coverage and also does not encompass the wavelength range of the broad Balmer absorption features, providing a cleaner tracer of the continuum. We then derived the monochromatic luminosity as $L_{\rm opt} = 4 \pi d^2 \, \nu F_\nu$, where $\nu$ is the central frequency of the ATLAS $o$-band and $F_\nu$ is the extinction-corrected flux density at that frequency.
We subtracted a constant companion contribution using the quiescent \gaia\ $G$-band photometry. Adopting $E(B-V)=0.27$ gives $A_G \simeq 0.64$ and a dereddened $G \simeq 19.06$, corresponding to $L_{\ast,\rm opt} \approx 1.6\times10^{33}$\,\lum\ in the ATLAS $o$-band (this should be regarded as an upper limit because any residual non-stellar quiescent light would reduce the stellar term).
We matched each companion-subtracted optical luminosity with an X-ray luminosity measurement within a temporal window of $\pm0.3$\,d (we adopted the 2--10\,keV band for the X-ray luminosity to enable direct comparison with the study by \citealt{Russell2006}). This procedure yielded a total of 10 paired data points. Over this sample, $L_{\ast,\rm opt} \lesssim 9-34\% L_{\rm opt}$ (median $\simeq$12\%).

\begin{figure}[!ht]
\includegraphics[width=0.45\textwidth]{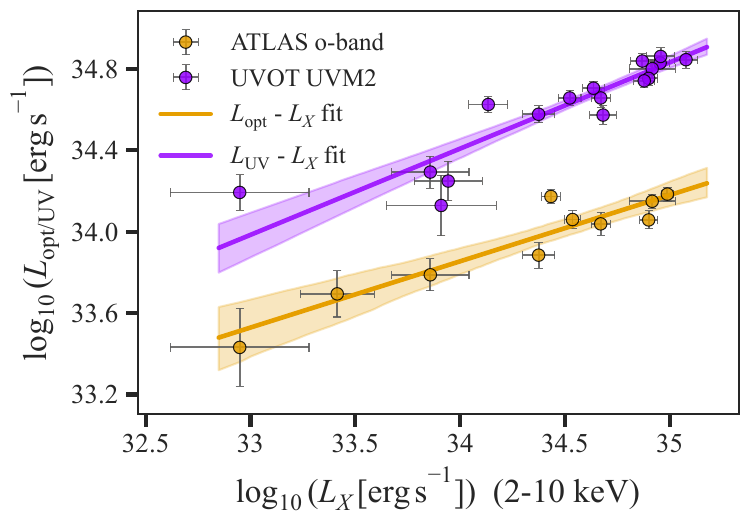}
\vspace{-0.3cm}
\caption{Correlation between ATLAS $o$-band or \swift/UVOT $uvm2$-band luminosities and quasi-simultaneous X-ray luminosities. The optical luminosities are corrected for a companion contribution estimated from the quiescent \gaia\ $G$-band flux (see Section\,\ref{sec:corr}). In some cases, error bars are smaller than the symbol size. The solid lines show the median posterior linear relations from the Bayesian regression; shaded regions indicate the 68\% credible intervals.}
\label{fig:lx_lopt}
\end{figure}

To account for both measurement uncertainties and intrinsic scatter, we performed a Bayesian linear regression \citep{Kelly2007}, sampling the posterior distributions of the slope ($\beta$), intercept ($\alpha$), and intrinsic random scatter around the best-fit model ($\sigma_\mathrm{int}$) for the relation: $\log_{10}\!\bigl(L_\mathrm{opt}/\mathrm{erg\,s^{-1}}\bigr)  = \alpha + \beta\, \log_{10}\!\bigl(L_X/\mathrm{erg\,s^{-1}}\bigr)$.
We ran the MCMC for at least 5000 iterations and, when needed, extended it up to 10000 iterations, checking convergence every 100 steps with the Gelman--Rubin diagnostic \citep{GelmanRubin1992} and requiring values below 1.1 for all monitored parameters. The resulting median values and 68\% credible intervals of the posterior distributions are:
$\beta_{\rm opt} = 0.32^{+0.07}_{-0.06}$, $\alpha_{\rm opt} = 23.1^{+2.1}_{-2.5}$, and $\sigma_\mathrm{int,opt} = 0.10^{+0.04}_{-0.03}~\mathrm{dex}$ (see Fig.\,\ref{fig:lx_lopt}).
Our measured slope is significantly lower than theoretical predictions for jet-dominated emission ($\beta_{\rm opt} \simeq 0.7$ for BH systems) and for X-ray reprocessing in the outer accretion disk ($\beta_{\rm opt} \simeq 0.5$), and slightly steeper than the nominal expectations for a viscously heated disk in a BH system ($\beta_{\rm opt} \approx 0.15$--0.25; \citealt{Russell2006} and refs. therein; see also \citealt{Coriat2009}). This disfavors a jet-dominated origin and instead points to predominantly disk emission during outburst, with viscous heating likely playing a major role and irradiation possibly providing an additional contribution.

We conducted a similar analysis for the UV emission, employing data from the \swift/UVOT $uvm2$ filter. After applying extinction corrections and computing monochromatic luminosities consistent with the optical analysis, we paired each $uvm2$ measurement with the strictly simultaneous \swift/XRT-derived X-ray luminosities. The Bayesian regression yielded posterior medians and 68\% credible intervals of $\beta_{\rm UV} = 0.44^{+0.07}_{-0.06}$, $\alpha_{\rm UV} = 19\pm2$, and $\sigma_\mathrm{int, UV} = 0.07^{+0.03}_{-0.02}~\mathrm{dex}$ (see Fig.\,\ref{fig:lx_lopt}).

Our analysis shows that the correlation slope is slightly steeper in the UV than in the optical, implying that shorter-wavelength emission responds more strongly to changes in $\dot{M}$. This trend is broadly consistent with an optically thick, viscously heated disk, for which $T(R)\propto \dot{M}^{1/4}R^{-3/4}$, so that progressively shorter wavelengths are preferentially emitted at smaller radii and higher temperatures. Since different bands trace different disk radii, different $L_\nu$–$\dot{M}$ scalings are therefore expected; in this framework, the larger $\beta_{\rm UV}$ indicates that the $uvm2$ emission is more tightly coupled to the hotter inner disk than the optical $o$ band. At the same time, the relatively large $\beta_{\rm UV}$ could also indicate a contribution from irradiation, which is expected to be comparatively stronger in the UV band.

  An additional clue to the origin of the optical outburst emission comes from the broad Balmer absorption troughs detected in the SALT spectra. Similar features have been reported in a number of LMXBs \citep[e.g.,][]{Callanan1995, Masetti1997, Soria2000, Elebert2009, Jimenez-Ibarra2019, Goodwin2020, Yao2021, Mereminskiy2022, Asquini2024, Corral-Santana2025}. A recent population study by \citet{MataSanchez2026} shows that such broad absorptions are common in LMXB outbursts, are stronger at shorter wavelengths, and exhibit nearly constant EW ratios across the population. In particular, our measured H$\beta$/H$\gamma$ absorption-EW ratios are fully consistent, within uncertainties, with the reported population value of $EW_{\rm abs}(H\beta)/EW_{\rm abs}(H\gamma)=1.05\pm0.05$. Together, these properties point to an origin in a stable, optically thick layer of the accretion disk beneath a hotter, chromosphere-like region. In this scenario, the absorption strength is further modulated by line filling and veiling from the X-ray reprocessed continuum. Within this framework, the blue continuum and Balmer absorption spectrum of \src\ are consistent with a disk-dominated origin of the optical emission during outburst. A detailed investigation of the line-formation geometry is beyond the scope of this work.

The short-timescale optical variability inferred from the SVOM/VT light curves is also consistent with BH LMXB phenomenology. On the native 100-s cadence, the measured VT$_{\rm R}$ fractional rms variability amplitude of $F_{\rm var}=2.2 \pm 0.1$\% is comparable to the few-per-cent optical values reported in BH systems from minute-cadence optical light curves \citep[e.g.,][]{Baglio2025,Saikia2026}, placing it well within the range typically seen in BH LMXBs. This modest fractional rms variability amplitude likely reflects low-amplitude intrinsic variability superposed on a dominant, steady disk continuum.

\subsection{Implications from Radio Non-Detections}
\label{sec:radio_discussion}
In X-ray binaries, radio emission serves as a key diagnostic of jet activity and offers crucial insights into the coupling between accretion inflow and outflows. Over the past two decades, empirical correlations have been established between the radio luminosity of compact jets ($L_r$) and the X-ray luminosity of the accretion flow ($L_X$) in both BH and NS systems, although with notably different normalizations and slopes (e.g., \citealt{Hannikainen1998,Corbel2003,Corbel2013,Gallo2003,Gallo2014,Gallo2018,Migliari2006,Tudor2017,vandenEijnden2021}). In particular, BH transients in the hard X-ray state typically follow a well-known non-linear relation of the form $L_r \propto L_X^\beta$, with a canonical slope of $\beta \simeq 0.6$, often referred to as the ``standard'' or ``radio-loud'' track.

\begin{figure*}[!ht]
\begin{center}
  \includegraphics[width=0.98\textwidth]{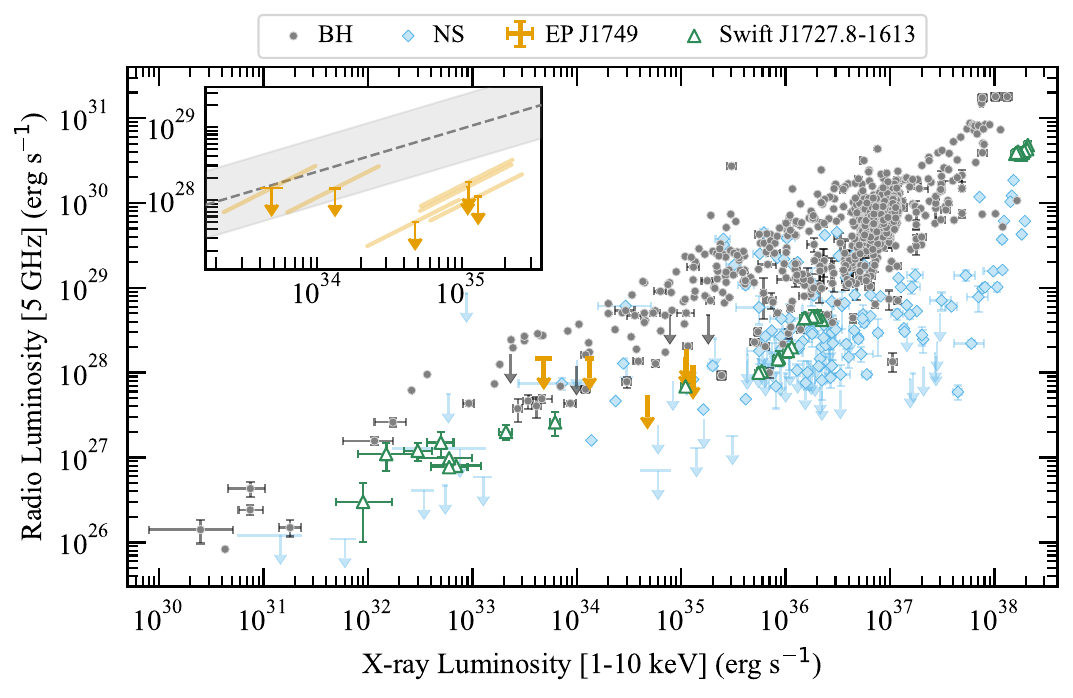}
 \vspace{-0.2cm}
\caption{Radio vs. X-ray luminosity for BH (black circles) and NS (cyan diamonds) LMXBs during the hard state.
\src\ is overplotted using orange squares (see Section\,\ref{sec:radio_discussion} for details), while Swift\,J1727.8$-$1613 is highlighted with green triangles for comparison. Detections are shown with error bars, while upper limits on the radio luminosity are indicated with downward-pointing caps. The inset displays only the measurements for \src. The orange diagonal bars represent the luminosity ranges accounting for the uncertainty on the distance. The gray dashed line and shaded region indicate the best-fit relation and corresponding 68\% credible interval for BH LMXBs (see \citealt{Gallo2018}).
}
\label{fig:radio_lx}
\end{center}
\end{figure*}

Our campaign provides a rare opportunity to place a newly discovered BH VFXT within this framework. Figure\,\ref{fig:radio_lx} shows the upper limits on the radio luminosity of \src, computed assuming a flat radio spectrum, against quasi-simultaneous X-ray luminosities spanning $\sim$1.4 orders of magnitude. For comparison, we include BH and NS LMXBs from the literature (for a compilation, see \citealt{Bahramian2022}). The radio non-detections place \src\ below the standard BH track, suggesting that compact jet emission is either intrinsically weak or strongly suppressed even at accretion luminosities where BH systems often show steady jets. An instructive comparison is provided by Swift\,J1727.8$-$1613: during its 2023--2024 outburst, this system followed a steeper relation ($\beta \gtrsim 1$) at higher luminosities ($L_X \gtrsim 10^{36}$\,\lum), then transitioned to a shallower one ($\beta \lesssim 0.3$) at lower X-ray luminosities, well below the standard BH track, before eventually rejoining the standard track (\citealt{Hughes2025}; see Fig.\,\ref{fig:radio_lx}). Although \src\ is constrained only by radio upper limits rather than detections, this case shows that large departures from the canonical BH $L_r$--$L_X$ relation do not by themselves rule out a BH accretor, but instead point to a broader diversity in jet--accretion coupling among hard-state BH LMXBs.

We emphasize that the location of \src\ in Fig.\,\ref{fig:radio_lx} remains the main source of tension with a BH interpretation, as the radio upper limits overlap the locus of some NS systems and fall below the standard BH $L_r$--$L_X$ relation, even when accounting for distance uncertainties. If the accretor were a NS, the low radio luminosity would be more naturally accommodated given the lower normalization of the NS correlation. At the same time, the behavior of Swift\,J1727.8$-$1613 discussed above shows that even confirmed BH systems can undergo large excursions below the canonical BH track, plausibly reflecting changes in jet production efficiency or jet--accretion coupling rather than a different accretor type. The present radio data alone therefore cannot discriminate between a radio-faint BH, a transiently suppressed or highly variable jet, and an NS accretor, and primarily indicate the absence of a bright quasi-steady jet during the sampled epochs. In contrast, several independent X-ray diagnostics favor a BH accretor: \src\ lies on the BH $\Gamma$--$L_X$ track, shows strong aperiodic variability characteristic of hard-state BHs, and exhibits no evidence for coherent pulsations or thermonuclear bursts. Taken together, these considerations suggest that \src\ is more plausibly an unusually radio-faint BH at these luminosities than a NS system.

\subsection{Multi‑episodic Outburst Behavior}

Based on the classification scheme proposed by \citet{Zhang2019}, \src\ appears to have undergone two full outbursts and one rebrightening during our observing campaign.

Several LMXBs have been observed to exhibit multi-episodic outbursts (e.g., \citealt{Cuneo2020,Dai2023,Nath2024,Manca2025}), in which the light curve departs from a single smooth peak and instead displays reflares, mini-outbursts, or consecutive activity episodes (see also \citealt{Tetarenko2016}). Multiple mechanisms have been proposed to account for this behaviour. Within the DIM, cooling fronts near the end of an outburst may stall or reverse, enabling renewed heating-front propagation and triggering secondary outbursts (e.g., \citealt{Hameury2020}). Irradiation and changes in the inner disk structure, such as truncation and refilling, can further shape this evolution. Alternatively, enhanced mass transfer from the companion -- potentially driven by irradiation -- can resupply the disk and reignite accretion. Mass loss through winds or outflows may additionally modulate the decay and produce rebrightenings.

The recurrence timescale of \src\ is broadly consistent with the viscous timescale at large radii in the accretion flow, i.e. the timescale on which matter or accretion-rate perturbations drift inward through the disk. For accretion flows around stellar-mass BHs, this timescale typically ranges from about a week to a couple of months for plausible values of viscosity ($\alpha \approx 0.1$), mass inflow scale height ($H/R \approx 0.1$--$0.3$), and outer disk radius ($R \approx 10^{11}$\,cm).
Moreover, the optical/UV–X-ray luminosity correlations of \src\ suggest that the variability is more consistent with disk-internal accretion-rate fluctuations than with a strongly irradiation-dominated optical response (see Section\,\ref{sec:corr}).
Overall, these findings are consistent with DIM-like ``reflares'', in which accretion is re-ignited on a viscous timescale as parts of the disk remain warm and ionized following the first detected event.

\section{Conclusions}
\label{sec:conclusions}
We presented the discovery and multi-wavelength characterization of the Galactic transient \src, detected by EP during a faint X-ray outburst in March 2025. Follow-up observations with multiple facilities revealed two major outbursts and a brief rebrightening over a $\sim$7-month period, during which the X-ray luminosity varied by more than three orders of magnitude. The temporal and spectral behavior of \src\ identifies it as a VFXT that remained in a low-luminosity, sub-Eddington state throughout its activity. Our main findings can be summarized as follows:

$\bullet$ The X-ray spectrum remained consistently hard ($\Gamma\simeq$1-2) throughout the outburst, was well described by thermal Comptonization of soft seed photons, and lacked any detectable thermal component. As the source faded, the spectrum slightly softened, matching the X-ray luminosity vs. spectral evolution seen in BH LMXBs accreting at very low rates and thus supporting a BH accretor. The X-ray emission exhibited strong aperiodic variability (rms $\simeq$30\%). The PDS displayed a break at a frequency of $\simeq$0.04\,Hz. Assuming this feature traces the viscous timescale in the inner flow, it would be consistent with a truncated accretion disk, with the inferred inner disk edge located at a few tens of gravitational radii (assuming a $\sim10\,M_\odot$ BH).

$\bullet$ The optical and UV counterparts brightened in tandem with the X-rays and exhibited a blue continuum with broad Balmer absorption features. Together with the optical/UV--X-ray correlations, this supports a disk-dominated origin of the optical/UV outburst emission, with viscous heating likely playing a major role and irradiation possibly contributing, especially in the UV.

$\bullet$ No radio counterpart was detected during our extensive monitoring campaign, placing stringent upper limits on jet activity during this outburst and in the largely unexplored luminosity regime of VFXTs.

Collectively, the observed properties support the classification of \src\ as a new VFXT BH candidate, although the current data do not yet allow for a definitive determination of the accretor type. Continued monitoring during future outbursts and in quiescence, particularly through deeper radio searches to probe jet activity and time-resolved optical spectroscopy to dynamically constrain the accretor mass, will be crucial to firmly establish its nature. If validated as a BH system, \src\ would add to the still small but rapidly growing population of BH VFXTs, offering a valuable probe of accretion and outflow physics at extremely low mass-transfer rates. Overall, this study underscores the ability of EP to uncover and characterize the faintest accreting compact objects in the Galaxy, opening new avenues for exploring mass accretion mechanisms at the lowest luminosities.

\begin{acknowledgments}
 We thank the referee for a constructive report that helped improve the clarity and focus of this work.

We thank the \nicer\ principal investigator, Keith Gendreau, for approving our Target of Opportunity (ToO) request and the operation team for executing the observations; we also thank the \nustar\ principal investigator, Fiona Harrison, for approving our Director's Discretionary Time (DDT) request and the operation team for executing the observations; we also thank the \swift\ deputy project scientist, Brad Cenko, and the \swift\ duty scientists and science planners, for making the \swift\ ToO observations possible. F.C.Z. thanks Alessandra Ambrifi for insightful discussions on the analysis of optical spectra.
This work is based on data obtained with: \emph{Einstein Probe}, a space mission supported by the Strategic Priority Program on Space Science of the Chinese Academy of Sciences, in collaboration with the European Space Agency, the Max Planck Institute for Extraterrestrial Physics (Germany), and the Centre National d'\'Etudes Spatiales (France); the \nicer\  mission, a 0.2--12\,keV X-ray telescope operating on the International Space Station; the \nustar\ mission, a project led by the California Institute of Technology, managed by the Jet Propulsion Laboratory, and funded by NASA; the \emph{Neil Gehrels Swift Observatory}, a NASA/UK/ASI mission; the Space Variable Objects Monitor (SVOM), a China--France joint mission led by the Chinese National Space Administration (CNSA, China), National Center for Space Studies (CNES, France) and the Chinese Academy of Sciences (CAS, China), dedicated to observing gamma-ray bursts and other transient phenomena in the energetic universe. VT was jointly developed by Xi'an Institute of Optics and Precision Mechanics (XIOPM), CAS and National Astronomical Observatories (NAOC), CAS. This research has made use of the \nustar\ Data Analysis Software (NuSTARDAS) jointly developed by the ASI Space Science Data Center (SSDC, Italy) and the California Institute of Technology (Caltech, USA). We also used software and tools provided by the High Energy Astrophysics Science Archive Research Center (HEASARC) Online Service.
This work has made use of data from the Asteroid Terrestrial-impact Last Alert System (ATLAS) project, which is primarily funded to search for near-Earth asteroids through NASA grants NN12AR55G, 80NSSC18K0284, and 80NSSC18K1575; byproducts of the NEO search include images and catalogs from the survey area. This work was partially funded by Kepler/K2 grant J1944/80NSSC19K0112 and HST GO-15889, and STFC grants ST/T000198/1 and ST/S006109/1. The ATLAS science products have been made possible through the contributions of the University of Hawaii Institute for Astronomy, the Queen’s University Belfast, the Space Telescope Science Institute, the South African Astronomical Observatory, and The Millennium Institute of Astrophysics (MAS), Chile.
This work is based on data products from VVV Survey observations made with the VISTA telescope at the ESO Paranal Observatory under programme ID 179.B-2002.
Some of the observations reported in this paper were obtained with the Southern African Large Telescope (SALT), as part of the Large Science Programme on transients (PI: D.~A.~H.~Buckley).
\texttt{IRAF} is distributed by the National Optical Astronomy Observatories, which are operated by the Association of Universities for Research in Astronomy, Inc., under a cooperative agreement with the NSF. The \texttt{IRAF} astronomical data reduction and analysis software package is available at \url{https://iraf-community.github.io}.
The MeerKAT telescope is operated by the South African Radio Astronomy Observatory (SARAO), which is a facility of the National Research Foundation, an agency of the Department of Science and Innovation. X-KAT is a large MeerKAT open-time programme to observe X-ray binaries in the radio band, performing weekly monitoring of bright, active systems, with capacity for higher cadence observations, and in coordination with large X-ray and optical monitoring programmes (MeerKAT Proposal ID: SCI-20230907-RF-01; PI: R.~P.~Fender).
We acknowledge the use of the Inter-University Institute for Data Intensive Astronomy (IDIA) data-intensive research cloud for data processing. IDIA is a South African university partnership involving the University of Cape Town, the University of Pretoria, and the University of the Western Cape. The Canadian Initiative for Radio Astronomy Data Analysis (CIRADA) is funded by a grant from the Canada Foundation for Innovation 2017 Innovation Fund (Project 35999), as well as by the Provinces of Ontario, British Columbia, Alberta, Manitoba and Quebec.

F.C.Z. is supported by a Ram\'on y Cajal fellowship (grant agreement RYC2021-030888-I). N.R. is supported by the European Research Council (ERC) via the Consolidator Grant ``MAGNESIA'' (No. 817661) and the Proof of Concept ``DeepSpacePulse'' (No. 101189496). F.C.Z., A.M., Y.L.W. and N.R. acknowledge support from the Spanish grant ID2023-153099NA-I00, and by the program Unidad de Excelencia Maria de Maeztu CEX2020-001058-M. Y.L.W. is supported by the China Scholarship Council (No. 202404910397). S.G. acknowledges the support of the CNES. S.E.M. acknowledges support from the INAF Fundamental Research Grant (2022) EJECTA. Y. X. acknowledges support from National Science Foundation of China through grants NSFC-12521005. Y.F.H. is supported by National Key R\&D Program of China (2021YFA0718500) and by the Xinjiang Tianchi Program. G.I. is supported by a Juan de la Cierva fellowship (JDC2024-053550-I). M.C.B. acknowledges support from the INAF Fundamental Research Grant XBOOM. R.P.F. acknowledges support from UK Research and Innovation, the European Research Council (ERC Synergy Grant `BlackHolistic', Reference 101071643) and the Hintze Family Charitable Foundation.
This work received financial support from INAF through the GRAWITA 2022 Large Program Grant.

\end{acknowledgments}




%
\facilities{\emph{Einstein Probe} (WXT and FXT), \nicer, \nustar, SVOM (MXT and VT), \swift\ (XRT and UVOT), ATLAS, SALT, MeerKAT, ADS, HEASARC.}
\software{
3D-NH-tool (\url{http://astro.uni-tuebingen.de/nh3d/nhtool});
Astrometry.net v0.98 \citep{Lang2010};
Astropy v7.2.0 \citep{astropy:2013,astropy:2018,astropy:2022};
Astroquery v0.4.11 \citep{Ginsburg2019};
CASA v6.7.3 \citep{casa};
CubiCal v1.6.4 \citep{cubical};
dust-extinction v1.7 \citep{Gordon2024};
dustmaps v1.0.14 \citep{Green2018};
FXTDAS v1.20 (\url{http://epfxt.ihep.ac.cn/analysis});
HEASoft v6.36.0 \citep{heasoft14};
HILIGT v1.7 \citep{Saxton2022,Konig2022};
LinMix (\url{https://github.com/jmeyers314/linmix});
Matplotlib v3.10.9 \citep{hunter07};
mw-plot v0.13.1;
MXT pipeline v1.14 \citep{Maggi2026};
NICERDAS v15;
nicerutil (\url{https://github.com/georgeyounes/NICERUTIL});
nDspec v0.2 Alpha \citep{Lucchini2025};
NumPy v2.4.4 \citep{harris20};
OxKAT \citep{oxkat};
pandas v3.0.2 \citep{mckinney2010};
PRESTO v5.2.0 \citep{Ransom2011};
PySALT v0.50 \citep{Crawford2010};
PyXspec v2.1.5 \citep{Gordon2021};
pyZDCF v1.0.3 \citep{Jankov2022};
relxill v2.8 \citep{Garcia2014};
SAOImageDS9 v8.6 \citep{joye03};
scikit-learn v1.8.0 \citep{Pedregosa2011};
SciPy v1.17.1 \citep{scipy20};
specutils v2.3.0 \citep{Earl2025};
Stingray v2.3.2 \citep{Huppenkothen2019,Bachetti2024};
Tricolour v0.2.0 \citep{tricolour};
XIMAGE v4.5.1 (\url{https://heasarc.gsfc.nasa.gov/docs/software/ximage/ximage.html});
XSPEC v12.15.1 \citep{Arnaud1996};
WebPIMMS (\url{https://heasarc.gsfc.nasa.gov/cgi-bin/Tools/w3pimms/w3pimms.pl});
WSClean v3.7 \citep{wsclean}.
}


\appendix
\restartappendixnumbering

\section{Log of Multi-band Observations}
This Appendix provides a log of the multi-wavelength observational campaign conducted on \src, along with archival X-ray observations of the field. Table\,\ref{tab:obsX} lists all pointed X-ray observations performed with \nicer, \swift/XRT, \nustar, and EP/FXT, along with the corresponding exposure times, count rates, photon indices, and both observed and unabsorbed fluxes. Tables\,\ref{tab:obsUV} and \ref{tab:obsVT} present the results from \swift/UVOT and SVOM/VT photometry, respectively. Table\,\ref{tab:obsmeerkat} summarizes the MeerKAT radio observations. Table\,\ref{tab:xobs_archive} reports the archival X-ray observations of the field performed with \ros, \xmm, and \emph{eROSITA}.

\begin{deluxetable*}{ccccccccc}
\tabletypesize{\tiny}
\setlength{\tabcolsep}{4.5pt}
\tablewidth{0pt}
\tablecaption{Log of pointed X-ray observations of \src\ along with spectral fitting results\label{tab:obsX}}
\tablehead{
  \colhead{X-ray Instrument\tablenotemark{a}} &
  \colhead{Obs.\ ID} &
  \colhead{Start (UTC)} &
  \colhead{Stop (UTC)} &
  \colhead{Exp.\ (ks)} &
  \colhead{Count rate\tablenotemark{b}} &
  \colhead{$\Gamma$} &
  \colhead{Obs./Unabs. Flux\tablenotemark{c}} &
  \colhead{$C$-stat (dof)}
}
\startdata
\nicer/XTI  &  8205360101    & 2025-03-26 13:22:40  &  2025-03-26 21:48:20  & 1.2  	& 13.4$\pm$0.1	     & 1.26$\pm$0.01 	& 6.7$\pm$0.1 /  $7.72^{+0.05}_{-0.15}$ & 365.81 (112) \\
\swift/XRT  &  00019649001 & 2025-03-30 15:48:23  & 2025-03-30 15:48:54  & 0.03   	& 0.4$\pm$0.1  	     & 1.8$\pm$0.6   	& 2$^{+2}_{-1}$ / $2.6^{+0.8}_{-0.7}$ & 5.30 (9)  \\
\nustar/FPMA+FPMB  & 91101307002 & 2025-03-30 22:06:32 & 2025-03-31 08:36:23 	& 21.5		& 1.243$\pm$0.008    & 1.75$\pm$0.01	& 2.88$^{+0.02}_{-0.03}$/ $4.23\pm0.07$ &   \\
SVOM/MXT    & 1426068190  &	2025-03-30 22:41:06  & 2025-03-31 06:29:50 & 8.4   	   & 0.18$\pm$0.01 	    & -- & $<$6.8/ $<$8.8 & 178.12 (170)   \\
EP/FXT      & 11900161664 & 2025-04-01 06:29:57 & 2025-04-01 06:55:16	& 1.5  		& 2.31$\pm$0.04 	& 1.70$\pm$0.03		& 3.0$\pm$0.1 / $3.50\pm0.09$ & 69.14 (68)  \\
\swift/XRT  &  00019649002 & 2025-04-02 16:46:40  & 2025-04-02 23:12:11  & 1.  		& 0.26$\pm$0.02 	& 1.8$\pm$0.1  		& 2.3$^{+0.1}_{-0.3}$ / $2.8\pm0.2$ & 158.82 (178)  \\
EP/FXT      & 11900166528 & 2025-04-04 08:11:46 & 2025-04-04 09:41:51	& 2.6 		& 1.94$\pm$0.03	      & 1.75$\pm$0.03	& 2.5$\pm$0.1 / $2.92\pm0.06$ & 110.15 (96)  \\
\swift/XRT  &  00019649003 & 2025-04-06 06:38:53  & 2025-04-06 06:51:53  & 0.8  	& 0.34$\pm$0.02 	& 1.6$\pm$0.1  		& 2.4$^{+0.1}_{-0.3}$ / $2.8\pm0.2$ & 144.27 (181)  \\
\nicer/XTI  &  8205360109    & 2025-04-08 08:06:00  &  2025-04-08 22:13:16  & 0.9  	& 8.1$\pm$0.1		& 1.46$\pm$0.02  	& 3.33$^{+0.09}_{-0.05}$ / $3.74^{+0.05}_{-0.06}$ & 132.23 (104) \\
\nicer/XTI  &  8205360110    & 2025-04-11 04:33:20  &  2025-04-11 11:02:20  & 0.4  	& 5.2$\pm$0.1		& 1.62$^{+0.04}_{-0.03}$  & 1.94$^{+0.08}_{-0.09}$ / $2.23^{+0.07}_{-0.05}$ & 81.15 (88) \\
\swift/XRT  &  00019649004 & 2025-04-13 09:03:17  & 2025-04-13 09:18:53  & 0.9  	& 0.20$\pm$0.01 	& 1.6$^{+0.2}_{-0.1}$  	& 1.4$^{+0.2}_{-0.1}$ / $1.7\pm0.1$ & 108.42 (137)  \\
\nicer/XTI  &  8205360111    & 2025-04-17 06:06:00  &  2025-04-17 06:22:20  & 0.5  	& 2.0$\pm$0.1		& 1.29$^{+0.06}_{-0.04}$ & 1.09$^{+0.1}_{-0.08}$ / $1.03^{+0.05}_{-0.04}$ & 46.33 (52) \\
\swift/XRT  &  00019649005 & 2025-04-20 04:59:18  & 2025-04-20 05:13:53  & 0.9  	& 0.037$\pm$0.007  & 2.7$\pm$0.4  			& 0.13$^{+0.03}_{-0.02}$ / $0.20^{+0.04}_{-0.03}$ & 16.48 (26)  \\
\swift/XRT  &  00019649006 & 2025-04-27 02:18:58  & 2025-04-27 02:31:53  & 0.8  	& $<$0.02   		& --   				& $<$0.06 / $<$0.09 & -- \\
\swift/XRT  &  00019649007 & 2025-05-04 13:51:30  & 2025-05-04 14:05:53  & 0.9  	& $<$0.009   		& --  	 			& $<$0.03 / $<$0.04 & -- \\
\swift/XRT  &  00019649008 & 2025-05-11 22:04:47  & 2025-05-11 22:21:52  & 1.0 		& 0.012$\pm$0.004  & 2.6$\pm$0.6  		& 0.04$^{+0.01}_{-0.02}$  / $0.07\pm0.02$ & 8.97 (9) \\
\swift/XRT  &  00019649009 & 2025-05-18 12:57:43  & 2025-05-18 13:14:55  & 0.9  	& $<$0.02  		    & --  			    & $<$0.06 / $<$0.09 & -- \\
EP/FXT      & 11900236160  & 2025-05-20 07:36:41 & 2025-05-20 09:37:33	& 3.5	 	& $<$0.01			& --				& $<$0.01 / $<$0.021 & --    \\
EP/FXT      & 11900239873  & 2025-05-23 09:24:49 & 2025-05-23 10:57:21	& 2.8	 	& $<$0.01   		& --				& $<$0.009 / $<$0.015 & --   \\
\swift/XRT  &  00019649010 & 2025-05-25 02:26:48  & 2025-05-25 02:41:53  & 0.9  	& $<$0.009 		    &  -- 			    & $<$0.03 / $<$0.04 & -- \\
EP/FXT      & 11900246528 & 2025-05-27 02:46:39 & 2025-05-27 03:26:13	& 2.4	 	& 0.027$\pm$0.004  	& 2.6$\pm$0.5		& 0.024$^{+0.007}_{-0.004}$ / $0.042^{+0.010}_{-0.009}$ & 0.07 (2)   \\
EP/FXT      & 11900251136 & 2025-05-30 03:07:08 & 2025-05-30 04:35:57	& 2.7	 	& 0.202$\pm$0.009  	& 2.0$\pm$0.1		& 0.20$\pm$0.02 / $0.24\pm0.02$ & 22.90 (22) 	  \\
\swift/XRT  &  00019649011 & 2025-06-01 21:39:51  & 2025-06-01 22:05:43  & 1.5  	& 0.052$\pm$0.006  	& 1.8$\pm$0.2   	& 0.45$^{+0.05}_{-0.07}$ /  $0.52\pm0.06$ & 44.60 (69) \\
EP/FXT      & 11900259712 & 2025-06-05 01:18:06 & 2025-06-05 01:51:35	& 2.0	 	& 0.71$\pm$0.02     & 1.8$\pm$0.1		& 0.81$^{+0.07}_{-0.04}$ / $1.03\pm0.04$ & 42.67 (30)  \\
\swift/XRT  &  00019649012 & 2025-06-08 14:27:14  & 2025-06-08 22:28:44  & 1.8   	& 0.17$\pm$0.01 	& 1.5$\pm$0.1  		& 2.6$^{+0.3}_{-0.2}$ / $2.9\pm0.2$ & 160.47 (207)  \\
\swift/XRT  &  00019649013 & 2025-06-15 00:45:55  & 2025-06-15 00:53:44  & 0.5 		& 0.36$\pm$0.03 	& 1.8$\pm$0.1  		& 3.0$^{+0.4}_{-0.3}$ / $3.6\pm0.3$ & 122.44 (118)  \\
\swift/XRT  &  00019649014 & 2025-06-18 09:31:59  & 2025-06-18 09:40:21  & 0.5 		& 0.42$\pm$0.03 	& 1.5$\pm$0.1  		& 3.5$^{+0.3}_{-0.4}$ / $3.9\pm0.3$ & 133.33 (159)  \\
EP/FXT      & 11900280576 & 2025-06-22 17:27:31 & 2025-06-22 20:59:09	& 3.6	  	& 2.52$\pm$0.03    	& 1.63$\pm$0.03		& 3.50$^{+0.04}_{-0.1}$ / $4.05\pm0.06$ & 158.17 (157)  \\
\swift/XRT  &  00019649016 & 2025-06-26 23:27:40  & 2025-06-26 23:34:47  & 0.4 		& 0.15$\pm$0.02 	& 1.9$\pm$0.2  		& 2.7$^{+0.6}_{-0.3}$ / $3.3\pm0.4$ & 39.97 (53)  \\
EP/FXT      & 11900289792 & 2025-06-29 12:35:17 & 2025-06-29 16:06:45	& 3.8		& 1.94$\pm$0.02    	& 1.63$\pm$0.03		& 2.65$^{+0.08}_{-0.09}$ / $3.07\pm0.05$ & 143.93 (132)  \\
\swift/XRT  &  00019649017 & 2025-07-01 04:29:19  & 2025-07-01 18:53:46  & 1.2 		& 0.32$\pm$0.02 	& 1.7$\pm$0.1  		& 2.4$\pm$0.2 / $2.9^{+0.2}_{-0.1}$ & 208.37 (224) \\
\swift/XRT  &  00019649018 & 2025-07-03 00:20:06  & 2025-07-03 17:37:44  & 1.2 		& 0.25$\pm$0.01 	& 1.5$\pm$0.1  		& 2.3$\pm$0.2 / $2.6^{+0.2}_{-0.1}$ & 149.30 (197)  \\
EP/FXT      & 11900298240 & 2025-07-06 02:41:04 & 2025-07-06 09:16:24	& 1.4		& 1.42$\pm$0.03     & 1.74$\pm$0.07		& 1.8$\pm$0.1 / $2.2\pm0.1$ & 45.10 (43) \\
\swift/XRT  &  00019649019 & 2025-07-11 19:59:02  & 2025-07-11 21:43:44  & 1.5 		& 0.23$\pm$0.01 	& 1.8$\pm$0.1  		& 1.47$^{+0.08}_{-0.10}$ / $1.8\pm0.1$ & 149.01 (207)  \\
EP/FXT      & 11900307584 & 2025-07-13 02:56:58 & 2025-07-13 06:25:19	& 2.6		& 0.94$\pm$0.02     & 1.82$\pm$0.07		& 1.10$\pm$0.07 / $1.32^{+0.07}_{-0.06}$ & 42.70 (48)   \\
\swift/XRT  &  00019649020 & 2025-07-17 06:52:05  & 2025-07-17 11:48:45  & 1.6  	& 0.17$\pm$0.01	   & 1.6$\pm$0.1 	 	& 1.3$^{+0.2}_{-0.1}$ / $1.4\pm0.1$ & 165.33 (188)  \\
\swift/XRT  &  00019649021 & 2025-07-22 12:51:13  & 2025-07-22 13:12:44  & 1.3  	& 0.19$\pm$0.01	   & 1.7$\pm$0.1 	 	& 1.03$\pm$0.08 / 1.2$\pm$0.08 & 132.82 (169)  \\
\swift/XRT  &  00019649022 & 2025-07-27 01:58:47  & 2025-07-28 02:57:43  & 1.3 		& 0.11$\pm$0.01	   & 1.7$\pm$0.1 	 	& 0.7$^{+0.11}_{-0.07}$ / $0.81^{+0.10}_{-0.06}$ & 118.22 (119)  \\
\swift/XRT  &  00019649023 & 2025-08-03 08:12:50  & 2025-08-03 11:25:44  & 0.9   	& 0.029$\pm$0.006	& 1.4$\pm$0.4 	 	& 0.24$\pm$0.07 / $0.26^{+0.09}_{-0.06}$ & 25.10 (24)  \\
\swift/XRT  &  00019649024 & 2025-08-10 02:00:30  & 2025-08-10 14:41:52  & 1.7   	& $<$0.01	       & -- 	 		    & $<$0.03 / $<$0.05 & -- \\
\swift/XRT  &  00019649025 & 2025-08-17 19:29:09  & 2025-08-17 19:42:54  & 0.8   	& 0.041$\pm$0.007	& 1.9$\pm$0.4 	 	& $0.24^{+0.05}_{-0.08}$ / $0.30^{+0.08}_{-0.06}$ & 37.44 (30)  \\
\swift/XRT  &  00019649026 & 2025-08-24 16:18:04  & 2025-08-24 16:26:53  & 0.5    	& 0.026$\pm$0.007	& 1.3$\pm$0.6 	 	& 0.24$^{+0.09}_{-0.10}$ / 0.26$^{+0.14}_{-0.09}$ & 15.87 (11)  \\
\swift/XRT\tablenotemark{d}  &  00019649028 & 2025-08-30 01:09:48  & 2025-08-30 21:52:53  & 1.7    	& 0.004$\pm$0.002	& 1.5$\pm$0.7 	 	& 0.03$\pm$0.01 / 0.04$^{+0.02}_{-0.01}$ &  10.60 (13) \\
\swift/XRT\tablenotemark{d}  &  00019649029 & 2025-09-07 00:42:06  & 2025-09-07 19:36:52  & 2.0    	& 0.003$\pm$0.001	& 1.5$\pm$0.7 	 	& 0.03$\pm$0.01 / 0.04$^{+0.02}_{-0.01}$ & 10.60 (13) \\
\swift/XRT\tablenotemark{e}  &  00019649030 & 2025-09-16 08:24:09 & 2025-09-16 22:35:53  & 2.1    	& $<$0.009   		& --  	 			& $<$0.03 / $<$0.04 & -- \\
\swift/XRT\tablenotemark{e}  &  00019649031 & 2025-09-21 05:18:39 & 2025-09-21 05:22:02  & 0.2  		& $<$0.009   		& --  	 			& $<$0.03 / $<$0.04 & -- \\
\enddata
\tablenotetext{a}{All \swift\ XRT observations were performed in PC mode; all EP observations were performed with both FXT detectors in FF mode.}
\tablenotetext{b}{The count rate is in the 0.5--10\,keV energy range, except for \nustar\ (3--40\,keV). Upper limits are quoted at 3$\sigma$ confidence level.}
\tablenotetext{c}{The flux is in the 0.5--10\,keV energy range. Upper limits are computed assuming an absorbed power law model with \nh $= 3.0\times10^{21}$\,cm$^{-2}$ and $\Gamma=2.5$, and are quoted at the 3$\sigma$ c.l.}
\tablenotetext{d}{These spectra were fitted together.}
\tablenotetext{e}{The datasets of these observations were co-added to improve sensitivity.}
\end{deluxetable*}

\begin{deluxetable*}{ccccc}
\tabletypesize{\small}
\tablewidth{0pt}
\tablecaption{Log of \swift/UVOT observations and photometry results\label{tab:obsUV}}
\tablehead{
  \colhead{Obs.\ ID} &
  \colhead{Start (UTC)} &
  \colhead{Stop (UTC)} &
  \colhead{Filter} &
  \colhead{Magnitude\tablenotemark{a}}
}
\startdata
00019649001 & 2025-03-30 15:48:23  & 2025-03-30 15:48:54	& uvm2 & $>$17.4 \\
00019649002 & 2025-04-02 16:46:44  & 2025-04-02 16:49:55  & uvm2 & 17.3$\pm$0.1 \\
00019649002 & 2025-04-02 22:59:12  & 2025-04-02 23:12:14  & uvm2 & 17.61$\pm$0.08 \\
00019649003 & 2025-04-06 06:38:53  & 2025-04-06 06:51:53  & uvw1 & 17.38$\pm$0.07 \\
00019649004 & 2025-04-13 09:03:17  & 2025-04-13 09:18:53  & uvm2 & 18.2$\pm$0.1 \\
00019649005 & 2025-04-20 04:59:18  & 2025-04-20 05:13:53  & uvw2 & 19.3$\pm$0.2 \\
00019649006 & 2025-04-27 02:18:58  & 2025-04-27 02:31:53  & u       & 	$>$20.0  \\
00019649007 & 2025-05-04 13:51:30  & 2025-05-04 14:05:53  & uvw1 & $>$19.8  \\
00019649008 & 2025-05-11 22:04:47  & 2025-05-11 22:21:52  & uvm2 & 19.2$\pm$0.2 \\
00019649009 & 2025-05-18 12:57:43  & 2025-05-18 13:14:55  & uvw2 & $>$19.9  \\
00019649010 & 2025-05-25 02:26:48  & 2025-05-25 02:41:53  & u    & 18.7$\pm$0.1 \\
00019649011 & 2025-06-01 21:39:51  & 2025-06-01 22:05:43  & uvm2 & 18.09$\pm$0.09 \\
00019649012 & 2025-06-08 14:27:14  & 2025-06-08 22:28:44  & uvm2 & 17.58$\pm$0.06 \\
00019649013 & 2025-06-15 00:45:55  & 2025-06-15 00:53:44  & uvm2 & 17.5$\pm$0.1 \\
00019649014 & 2025-06-18 09:31:59  & 2025-06-18 09:40:21  & uvm2 & 17.5$\pm$0.1 \\
00019649016 & 2025-06-26 23:27:40  & 2025-06-26 23:34:47  & uvm2 & 17.65$\pm$0.11 \\
00019649017 & 2025-07-01 04:29:19  & 2025-07-01 18:53:46  & uvm2 & 17.77$\pm$0.08 \\
00019649018 & 2025-07-03 00:20:06  & 2025-07-03 17:37:44  & uvm2 & 17.80$\pm$0.08 \\
00019649019 & 2025-07-11 19:59:02  & 2025-07-11 21:43:44  & uvm2 & 18.0$\pm$0.1 \\
00019649020 & 2025-07-17 06:52:05  & 2025-07-17 11:48:45  & uvm2 & 17.89$\pm$0.08 \\
00019649021 & 2025-07-22 12:51:13  & 2025-07-22 13:12:44  & uvm2 & 18.01$\pm$0.09 \\
00019649022 & 2025-07-27 01:58:47  & 2025-07-28 02:57:43  & uvm2 & 18.2$\pm$0.1 \\
00019649023 & 2025-08-03 08:12:50  & 2025-08-03 11:25:44  & uvm2 & 19.0$\pm$0.2 \\
00019649024 & 2025-08-10 02:00:30  & 2025-08-10 14:41:52  & uvm2 & $>$19.9 \\
00019649025 & 2025-08-17 19:29:09  & 2025-08-17 19:42:54  & uvm2 & 18.9$\pm$0.2 \\
00019649026 & 2025-08-24 16:18:04  & 2025-08-24 16:26:53  & uvm2 & 19.3$\pm$0.3 \\
00019649028\tablenotemark{b1} & 2025-08-30 01:09:48  & 2025-08-30 21:52:53  & uvm2 & $>$20.4 \\
00019649029\tablenotemark{b1} & 2025-09-07 00:42:06  & 2025-09-07 19:36:52  & uvm2 & $>$20.4 \\
00019649030\tablenotemark{b2} & 2025-09-16 08:24:09  & 2025-09-16 22:35:53   & uvm2 & $>$20.1 \\
00019649031\tablenotemark{b2} & 2025-09-21 05:18:39  & 2025-09-21 05:22:02   & uvm2 & $>$20.1 \\
\enddata
\tablenotetext{a}{All magnitudes are in the Vega system. Reported values are either detections with 1$\sigma$ uncertainties (combining statistical and systematic errors in quadrature) or 3$\sigma$ upper limits.}
\tablenotetext{b}{Images from these datasets were co-added to improve sensitivity.}
\end{deluxetable*}

\begin{deluxetable*}{cccc}
\tabletypesize{\tiny}
\tablewidth{0pt}
\tablecaption{Log of SVOM/VT observations and photometry results\label{tab:obsVT}}
\tablehead{
  \colhead{Start (UTC)} &
  \colhead{Stop (UTC)} &
  \colhead{Band} &
  \colhead{Magnitude (AB)}
}
\startdata
2025-03-30 22:44:11 & 2025-03-30 22:45:51 & $VT_{\rm B}$, $VT_{\rm R}$ & 17.19$\pm$0.02, 17.09$\pm$0.02 \\
2025-03-30 22:45:51 & 2025-03-30 22:47:31 & $VT_{\rm B}$, $VT_{\rm R}$ & 17.21$\pm$0.02, 17.07$\pm$0.02 \\
2025-03-30 22:47:31 & 2025-03-30 22:49:11 & $VT_{\rm B}$, $VT_{\rm R}$ & 17.19$\pm$0.02, 17.04$\pm$0.02 \\
2025-03-30 23:51:14 & 2025-03-30 23:52:54 & $VT_{\rm B}$, $VT_{\rm R}$ & 17.21$\pm$0.02, 17.11$\pm$0.02 \\
2025-03-30 23:52:54 & 2025-03-30 23:54:34 & $VT_{\rm B}$, $VT_{\rm R}$ & 17.20$\pm$0.02, 17.05$\pm$0.02 \\
2025-03-30 23:54:34 & 2025-03-30 23:56:14 & $VT_{\rm B}$, $VT_{\rm R}$ & 17.20$\pm$0.02, 17.12$\pm$0.02 \\
2025-03-30 23:56:14 & 2025-03-30 23:57:54 & $VT_{\rm B}$, $VT_{\rm R}$ & 17.21$\pm$0.02, 17.05$\pm$0.02 \\
2025-03-30 23:57:54 & 2025-03-30 23:59:34 & $VT_{\rm B}$, $VT_{\rm R}$ & 17.22$\pm$0.02, 17.10$\pm$0.02 \\
2025-03-30 23:59:34 & 2025-03-31 00:01:14 & $VT_{\rm B}$, $VT_{\rm R}$ & 17.23$\pm$0.02, 17.06$\pm$0.02 \\
2025-03-31 00:01:14 & 2025-03-31 00:02:54 & $VT_{\rm B}$, $VT_{\rm R}$ & 17.20$\pm$0.02, 17.10$\pm$0.02 \\
2025-03-31 00:02:54 & 2025-03-31 00:04:34 & $VT_{\rm B}$, $VT_{\rm R}$ & 17.23$\pm$0.02, 17.07$\pm$0.02 \\
2025-03-31 00:04:34 & 2025-03-31 00:06:14 & $VT_{\rm B}$, $VT_{\rm R}$ & 17.23$\pm$0.02, 17.06$\pm$0.02 \\
2025-03-31 00:06:14 & 2025-03-31 00:07:54 & $VT_{\rm B}$, $VT_{\rm R}$ & 17.20$\pm$0.02, 17.08$\pm$0.02 \\
2025-03-31 00:07:54 & 2025-03-31 00:09:34 & $VT_{\rm B}$, $VT_{\rm R}$ & 17.22$\pm$0.02, 17.06$\pm$0.02 \\
2025-03-31 00:09:34 & 2025-03-31 00:11:14 & $VT_{\rm B}$, $VT_{\rm R}$ & 17.23$\pm$0.02, 17.13$\pm$0.02 \\
2025-03-31 00:11:14 & 2025-03-31 00:12:54 & $VT_{\rm B}$, $VT_{\rm R}$ & 17.22$\pm$0.02, 17.08$\pm$0.02 \\
2025-03-31 00:12:54 & 2025-03-31 00:14:34 & $VT_{\rm B}$, $VT_{\rm R}$ & 17.20$\pm$0.02, 17.05$\pm$0.02 \\
2025-03-31 00:14:34 & 2025-03-31 00:16:14 & $VT_{\rm B}$, $VT_{\rm R}$ & 17.23$\pm$0.02, 17.08$\pm$0.02 \\
2025-03-31 00:16:14 & 2025-03-31 00:17:54 & $VT_{\rm B}$, $VT_{\rm R}$ & 17.22$\pm$0.02, 17.07$\pm$0.02 \\
2025-03-31 00:17:54 & 2025-03-31 00:19:34 & $VT_{\rm B}$, $VT_{\rm R}$ & 17.22$\pm$0.02, 17.07$\pm$0.01 \\
2025-03-31 00:19:34 & 2025-03-31 00:21:14 & $VT_{\rm B}$, $VT_{\rm R}$ & 17.21$\pm$0.02, 17.07$\pm$0.02 \\
2025-03-31 00:21:14 & 2025-03-31 00:22:54 & $VT_{\rm B}$, $VT_{\rm R}$ & 17.23$\pm$0.02, 17.09$\pm$0.02 \\
2025-03-31 00:22:54 & 2025-03-31 00:24:34 & $VT_{\rm B}$, $VT_{\rm R}$ & 17.23$\pm$0.02, 17.06$\pm$0.02 \\
2025-03-31 00:24:34 & 2025-03-31 00:26:14 & $VT_{\rm B}$, $VT_{\rm R}$ & 17.23$\pm$0.02, 17.10$\pm$0.02 \\
2025-03-31 01:26:39 & 2025-03-31 01:28:19 & $VT_{\rm B}$, $VT_{\rm R}$ & 17.21$\pm$0.02, 17.10$\pm$0.02 \\
2025-03-31 01:28:19 & 2025-03-31 01:29:59 & $VT_{\rm B}$, $VT_{\rm R}$ & 17.22$\pm$0.02, 17.14$\pm$0.02 \\
2025-03-31 01:29:59 & 2025-03-31 01:31:39 & $VT_{\rm B}$, $VT_{\rm R}$ & 17.23$\pm$0.02, 17.16$\pm$0.02 \\
2025-03-31 01:31:39 & 2025-03-31 01:33:19 & $VT_{\rm B}$, $VT_{\rm R}$ & 17.22$\pm$0.02, 17.15$\pm$0.02 \\
2025-03-31 01:33:19 & 2025-03-31 01:34:59 & $VT_{\rm B}$, $VT_{\rm R}$ & 17.22$\pm$0.02, 17.11$\pm$0.02 \\
2025-03-31 01:34:59 & 2025-03-31 01:36:39 & $VT_{\rm B}$, $VT_{\rm R}$ & 17.22$\pm$0.02, 17.14$\pm$0.03 \\
2025-03-31 01:36:39 & 2025-03-31 01:38:19 & $VT_{\rm B}$, $VT_{\rm R}$ & 17.21$\pm$0.02, 17.13$\pm$0.02 \\
2025-03-31 01:38:19 & 2025-03-31 01:39:59 & $VT_{\rm B}$, $VT_{\rm R}$ & 17.24$\pm$0.02, 17.13$\pm$0.02 \\
2025-03-31 01:39:59 & 2025-03-31 01:41:39 & $VT_{\rm B}$, $VT_{\rm R}$ & 17.25$\pm$0.02, 17.17$\pm$0.03 \\
2025-03-31 01:41:39 & 2025-03-31 01:43:19 & $VT_{\rm B}$, $VT_{\rm R}$ & 17.23$\pm$0.02, 17.11$\pm$0.02 \\
2025-03-31 01:43:19 & 2025-03-31 01:44:59 & $VT_{\rm B}$, $VT_{\rm R}$ & 17.24$\pm$0.02, 17.11$\pm$0.02 \\
2025-03-31 01:44:59 & 2025-03-31 01:46:39 & $VT_{\rm B}$, $VT_{\rm R}$ & 17.23$\pm$0.02, 17.07$\pm$0.02 \\
2025-03-31 01:46:39 & 2025-03-31 01:48:19 & $VT_{\rm B}$, $VT_{\rm R}$ & 17.20$\pm$0.02, 17.11$\pm$0.02 \\
2025-03-31 01:48:19 & 2025-03-31 01:49:59 & $VT_{\rm B}$, $VT_{\rm R}$ & 17.21$\pm$0.02, 17.11$\pm$0.02 \\
2025-03-31 01:49:59 & 2025-03-31 01:51:39 & $VT_{\rm B}$, $VT_{\rm R}$ & 17.22$\pm$0.02, 17.08$\pm$0.01 \\
2025-03-31 01:51:39 & 2025-03-31 01:53:19 & $VT_{\rm B}$, $VT_{\rm R}$ & 17.23$\pm$0.02, 17.13$\pm$0.02 \\
2025-03-31 01:54:59 & 2025-03-31 01:56:39 & $VT_{\rm B}$, $VT_{\rm R}$ & 17.20$\pm$0.02, 17.08$\pm$0.02 \\
2025-03-31 01:56:39 & 2025-03-31 01:58:19 & $VT_{\rm B}$, $VT_{\rm R}$ & 17.22$\pm$0.02, 17.14$\pm$0.02 \\
2025-03-31 01:58:19 & 2025-03-31 01:59:59 & $VT_{\rm B}$, $VT_{\rm R}$ & 17.21$\pm$0.02, 17.07$\pm$0.02 \\
2025-03-31 01:59:59 & 2025-03-31 02:01:39 & $VT_{\rm B}$, $VT_{\rm R}$ & 17.21$\pm$0.02, 17.07$\pm$0.02 \\
2025-03-31 02:01:39 & 2025-03-31 02:03:19 & $VT_{\rm B}$, $VT_{\rm R}$ & 17.21$\pm$0.02, 17.10$\pm$0.02 \\
2025-03-31 03:05:23 & 2025-03-31 03:07:03 & $VT_{\rm B}$, $VT_{\rm R}$ & 17.22$\pm$0.02, 17.06$\pm$0.02 \\
2025-03-31 03:07:03 & 2025-03-31 03:08:43 & $VT_{\rm B}$, $VT_{\rm R}$ & 17.19$\pm$0.02, 17.05$\pm$0.01 \\
2025-03-31 03:08:43 & 2025-03-31 03:10:23 & $VT_{\rm B}$, $VT_{\rm R}$ & 17.22$\pm$0.02, 17.11$\pm$0.02 \\
2025-03-31 03:10:23 & 2025-03-31 03:12:03 & $VT_{\rm B}$, $VT_{\rm R}$ & 17.24$\pm$0.02, 17.10$\pm$0.02 \\
2025-03-31 03:12:03 & 2025-03-31 03:13:43 & $VT_{\rm B}$, $VT_{\rm R}$ & 17.22$\pm$0.02, 17.07$\pm$0.02 \\
2025-03-31 03:34:34 & 2025-03-31 03:36:14 & $VT_{\rm B}$ & 17.23$\pm$0.02 \\
2025-03-31 03:36:14 & 2025-03-31 03:37:54 & $VT_{\rm B}$, $VT_{\rm R}$ & 17.24$\pm$0.02, 17.12$\pm$0.02 \\
2025-03-31 03:37:54 & 2025-03-31 03:39:34 & $VT_{\rm B}$, $VT_{\rm R}$ & 17.20$\pm$0.02, 17.05$\pm$0.02 \\
2025-03-31 03:39:34 & 2025-03-31 03:41:14 & $VT_{\rm B}$, $VT_{\rm R}$ & 17.22$\pm$0.02, 17.10$\pm$0.02 \\
2025-03-31 04:40:48 & 2025-03-31 04:42:28 & $VT_{\rm B}$, $VT_{\rm R}$ & 17.23$\pm$0.02, 17.11$\pm$0.02 \\
2025-03-31 04:42:28 & 2025-03-31 04:44:08 & $VT_{\rm B}$, $VT_{\rm R}$ & 17.23$\pm$0.02, 17.11$\pm$0.02 \\
2025-03-31 04:44:08 & 2025-03-31 04:45:48 & $VT_{\rm B}$, $VT_{\rm R}$ & 17.21$\pm$0.02, 17.14$\pm$0.02 \\
2025-03-31 04:45:48 & 2025-03-31 04:47:28 & $VT_{\rm B}$, $VT_{\rm R}$ & 17.20$\pm$0.02, 17.06$\pm$0.02 \\
2025-03-31 04:47:28 & 2025-03-31 04:49:08 & $VT_{\rm B}$, $VT_{\rm R}$ & 17.24$\pm$0.02, 17.10$\pm$0.02 \\
2025-03-31 04:49:08 & 2025-03-31 04:50:48 & $VT_{\rm B}$, $VT_{\rm R}$ & 17.24$\pm$0.02, 17.09$\pm$0.02 \\
2025-03-31 04:50:48 & 2025-03-31 04:52:28 & $VT_{\rm B}$, $VT_{\rm R}$ & 17.25$\pm$0.02, 17.10$\pm$0.02 \\
2025-03-31 06:17:53 & 2025-03-31 06:19:33 & $VT_{\rm B}$, $VT_{\rm R}$ & 17.24$\pm$0.03, 17.15$\pm$0.02 \\
2025-03-31 06:19:33 & 2025-03-31 06:21:13 & $VT_{\rm B}$, $VT_{\rm R}$ & 17.21$\pm$0.02, 17.06$\pm$0.02 \\
2025-03-31 06:21:13 & 2025-03-31 06:22:53 & $VT_{\rm B}$, $VT_{\rm R}$ & 17.20$\pm$0.02, 17.06$\pm$0.02 \\
2025-03-31 06:22:53 & 2025-03-31 06:24:33 & $VT_{\rm B}$, $VT_{\rm R}$ & 17.21$\pm$0.02, 17.08$\pm$0.02 \\
2025-03-31 06:24:33 & 2025-03-31 06:26:13 & $VT_{\rm B}$, $VT_{\rm R}$ & 17.23$\pm$0.02, 17.08$\pm$0.02 \\
2025-03-31 06:26:13 & 2025-03-31 06:27:53 & $VT_{\rm B}$, $VT_{\rm R}$ & 17.23$\pm$0.02, 17.12$\pm$0.02 \\
2025-03-31 06:27:53 & 2025-03-31 06:29:33 & $VT_{\rm B}$, $VT_{\rm R}$ & 17.24$\pm$0.02, 17.07$\pm$0.02 \\
2025-03-31 06:29:33 & 2025-03-31 06:31:13 & $VT_{\rm B}$, $VT_{\rm R}$ & 17.24$\pm$0.02, 17.12$\pm$0.02 \\
\enddata
\end{deluxetable*}

\begin{deluxetable*}{ccccc}
\tabletypesize{\small}
\tablewidth{0pt}
\tablecaption{Log of MeerKAT observations and 3$\sigma$ upper limits on the flux density in the L-band. All observations lasted 15\,min.}
\label{tab:obsmeerkat}
\tablehead{
  \colhead{Epoch} &
  \colhead{Obs.\ ID} &
  \colhead{Observation start date} &
  \colhead{Flux density} &
  \colhead{Flux density [stacked]} \\
  \colhead{} &
  \colhead{} &
  \colhead{YYYY-MM-DD hh:mm:ss (UTC)} &
  \colhead{($\mu$Jy)} &
   \colhead{($\mu$Jy)}
}
\startdata
1 & 1743901392 & 2025-04-06 01:04:42  & $<57.9$  & $<62.0$ \\
2 & 1744418777 & 2025-04-12 00:47:47  & $<66.4$  &   \\  \hline
3 & 1745184398 & 2025-04-20 21:28:06  & $<68.4$  &   \\ \hline
4 & 1745797469 & 2025-04-27 23:45:57  & $<71.6$  & $<36.3$  \\
5 & 1746313275 & 2025-05-03 23:02:43  & $<77.8$  &   \\
6 & 1746915367 & 2025-05-10 22:17:34  & $<73.0$  &   \\
7 & 1747597679 & 2025-05-18 19:49:39  & $<79.4$  &   \\
8 & 1747944679 & 2025-05-22 20:11:52  & $<75.6$  &   \\ \hline
9 & 1748721677 & 2025-05-31 20:02:46  & $<67.8$  &   \\
10 & 1749501064 & 2025-06-09 20:31:37  & $<69.6$  & \\  \hline
11 & 1750009695 & 2025-06-15 19:43:51  & $<73.7$  & $<48.3$  \\
12 & 1750612578 & 2025-06-22 17:17:46  & $<64.8$  &   \\
13 & 1751216713 & 2025-06-29 17:06:42  & $<67.4$  &   \\  \hline
14 & 1751822175 & 2025-07-06 17:17:43  & $<71.3$  & $<18.9$  \\
15 & 1752252335 & 2025-07-11 16:47:03  & $<58.9$  &   \\
16 & 1752852856 & 2025-07-18 15:37:09  & $<63.3$  &   \\
17 & 1753636519 & 2025-07-27 17:16:47  & $<57.9$  &   \\
\enddata
\end{deluxetable*}

\begin{deluxetable}{lcccc}
\tablecaption{Archival observations of \src\ in the soft X-ray band
\label{tab:xobs_archive}}
\tablehead{
\colhead{Mission} & \colhead{Observation Date} & \colhead{Exp.} &
\colhead{Count rate\tablenotemark{a}} & \colhead{Obs./Unabs. Flux\tablenotemark{b}} \\
\colhead{} & \colhead{YYYY-MM-DD} &  \colhead{(s)} & \colhead{(counts\,s$^{-1}$)} & \colhead{($10^{-12}$\,\flux)}
}
\startdata
\ros\       & 1990-08-30   & 219  & $<0.06$  & $<1.3$ / $<1.9$ \\
\hline
\xmm\       & 2006-09-20   & 30   & $<0.25$  & $<0.6$ / $<0.8$ \\
            & 2015-09-28   & 7    & $<1.07$  & $<2.3$ / $<3.5$ \\
            & 2018-09-07   & 7    & $<1.26$  & $<2.8$ / $<4.1$ \\
\hline
\emph{eROSITA}     & 2020-03-30   & 105  & $<0.14$  & $<0.2$ / $<0.3$ \\
\enddata
\tablenotetext{a}{Count rates refer to the 0.2--2\,keV band for \ros\ and the 0.2--12\,keV band for \xmm\ and \emph{eROSITA}.}
\tablenotetext{b}{Flux upper limits are quoted at the 3$\sigma$ c.l., and are computed over the 0.5--10\,keV energy range assuming an absorbed power law model with \nh $= 3.0\times10^{21}$\,cm$^{-2}$ and $\Gamma=2.5$.}
\end{deluxetable}

\clearpage

\section{Gaia Counterpart Identification}
\label{app:lr_method}

Within the \swift\ error circle, we identify two stars listed in \gaia\ DR3 \citep{GaiaDR32023}. The closest, \gaia\ DR3 4036709435120357888, has a magnitude of $G \simeq 19.7$\,mag, a color index of BP--RP $= 1.06$\,mag, and lies $0.47\arcsec$ from the nominal \swift\ position. The second, \gaia\ DR3 4036709435120359168, has a magnitude of $G \simeq 20.2$\,mag  (with no available color information) and is located $1.54\arcsec$ away.

We quantified the probability of association using the likelihood-ratio (LR) method described by \citet{Sutherland1992}, which compares the probability that a given source is the true counterpart against the probability that it is an unrelated field object. Specifically, we define

\begin{equation}
LR = \frac{q(m) \, f(r)}{n(m)},
\end{equation}

where $q(m)$ is the expected distribution of the optical counterparts as a function of magnitude, $f(r)$ is the two-dimensional Gaussian probability density of the positional offset $r$ between the X-ray and optical positions, and $n(m)$ is the local surface density of \gaia\ sources of magnitude $m$.
We estimate $n(m)$ using an annular region between $5\arcmin$ and $15\arcmin$ from the X-ray position --- broad enough to avoid contamination from any local overdensity of unrelated sources, yet close enough to sample the same stellar population and extinction as the target field.
The reliability of candidate $i$ being the correct counterpart is then given by
\begin{equation}
\mathcal{R}_i = \frac{LR_i}{\sum_j LR_j + (1 - Q)} ,
\end{equation}
where the sum runs over the two identified candidates for \src, and
$Q$ is the probability that the optical counterpart of the X-ray source is brighter than the \gaia\ magnitude limit ($G \lesssim 20.7$\,mag\footnote{See \url{https://www.cosmos.esa.int/web/gaia/science-performance}}).
Assuming $Q = 0.9$, we find $LR = 2.26$ and $\mathcal{R} = 0.81$ for the nearest \gaia\ source, and $LR = 0.42$ and $\mathcal{R} = 0.15$ for the other \gaia\ source. Varying $Q$ between 0.5 and 1.0 changes the nearest-source reliability from 0.63 to 0.84. These findings point to \gaia\ DR3 4036709435120357888 as the optical counterpart of \src.

\section{Archival Radio Imaging}
\label{sec:radio_cutouts}

\begin{figure*}[!ht]
\begin{center}
\includegraphics[width=0.8\textwidth]{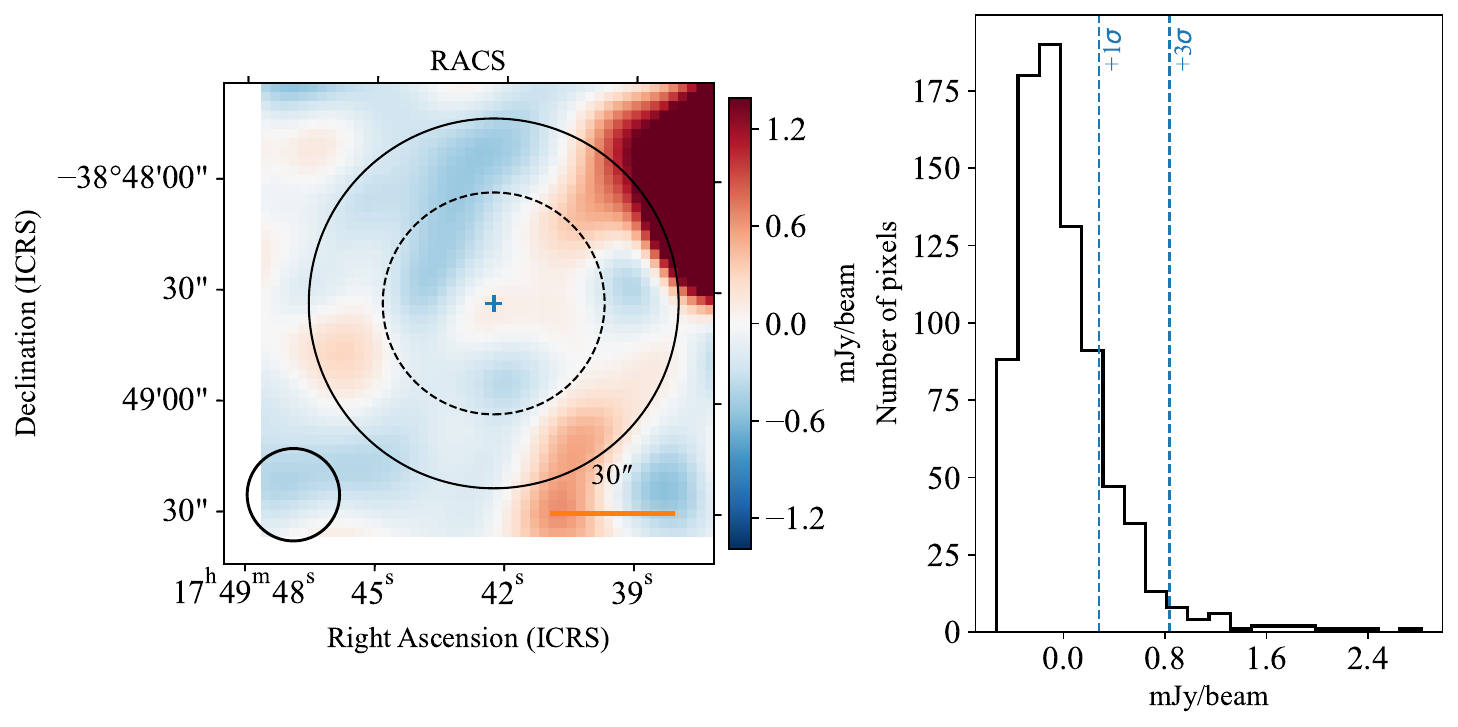}
\includegraphics[width=0.75\textwidth]{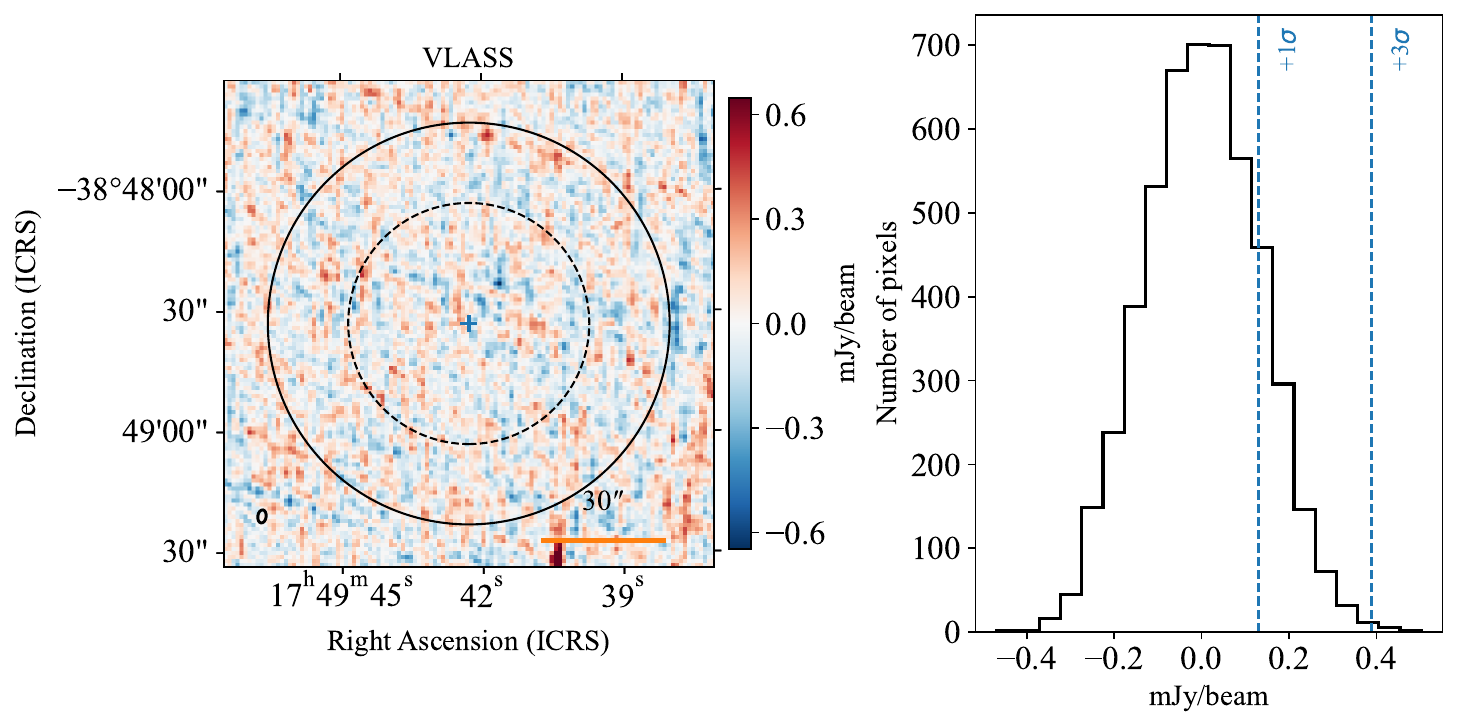}
\vspace{-0.2cm}
\caption{\emph{Top-left}: 1\arcmin$\times$1\arcmin\ radio Stokes-I intensity maps from the RACS survey, shown in units of mJy/beam (central frequency of 887.5\,MHz). The blue cross marks the target position. The dashed and solid black circles denote the inner and outer boundaries of the annular region (30\arcsec--50\arcsec) centered on the source position used for noise estimation. The beam size is shown as a black ellipse in the bottom-left. \emph{Top-right}: Histogram of pixel values within the annulus. Vertical dashed lines mark $+$1$\sigma$ and $+$3$\sigma$ from the median. \emph{Bottom}: Same layout as above, but using a cutout from the VLASS survey (central frequency of 3\,GHz).}
\label{fig:radio_cutouts}
\end{center}
\end{figure*}

We searched for a radio counterpart to \src\ in the Rapid ASKAP Continuum Survey (RACS; central frequency of 887.5\,MHz; \citealt{McConnell2020, Hale2021}) and the Karl G. Jansky Very Large Array Sky Survey (VLASS; 3\,GHz; \citealt{Lacy2020}). We retrieved 1\arcmin-radius Stokes-I image cutouts centered on the source position from the RACS-DR1 data via the CSIRO ASKAP Science Data Archive (CASDA\footnote{\url{https://research.csiro.au/casda/}}), and from VLASS via the Canadian Astronomy Data Centre (CADC\footnote{\url{https://www.cadc-ccda.hia-iha.nrc-cnrc.gc.ca/en/vlass/}}). In particular, the position of \src\ was covered using the VLA on 2019 July 5, 2022 February 14 and 2024 September 21. No emission was detected at the position of \src\ in either RACS or VLASS images (see Fig.\,\ref{fig:radio_cutouts}). We estimated the local rms background noise using an annular region with inner and outer radii of 30\arcsec\ and 50\arcsec, respectively. We then computed $3\sigma$ upper limits on the brightness of any radio emission at the source position as 3$\times$rms noise, obtaining $\simeq$0.8\,mJy\,beam$^{-1}$ at 887.5\,MHz and $\simeq$0.4\,mJy\,beam$^{-1}$ at 3\,GHz.

\section{X-ray Timing and Burst Search Methodology}
\label{sec:X_searches}
We searched for periodic X-ray signals using the \nicer\ and \nustar\ datasets, given their superior time resolution and count statistics.

We first applied the method developed by \citet{Israel1996}, which is optimized for detecting periodic signals in the presence of red noise in the power spectrum. In short, we computed the Leahy-normalized PDS from the event files and applied an ad hoc smoothing algorithm to estimate the local continuum around each frequency bin.
The raw PDS was then divided by the smoothed continuum to produce a whitened spectrum. No statistically significant peaks were detected in this spectrum above a 3.5$\sigma$ threshold, accounting for the number of independent frequency trials. The most stringent upper limit on the pulsed fraction was obtained from the \nustar\ data: $<$6\% over the 1--1000\,Hz frequency range (3$\sigma$; assuming a sinusoidal signal).

We conducted complementary searches using only the \nustar\ data. First, we employed a Fourier-domain acceleration search algorithm to account for potential Doppler smearing of a putative pulsar signal caused by binary orbital motion. We applied a Fourier transform to the photon arrival times and subtracted red noise to flatten the low-frequency baseline. We then performed an acceleration search on 2\,ks-long time segments using the \texttt{accelsearch} routine from the \texttt{PRESTO} software package \citep{Ransom2002,Ransom2011}. We allowed the highest harmonic of any signal to drift by up to 200 Fourier bins, corresponding to a maximum line-of-sight acceleration of $a_{\rm max} \approx 15 P_{\rm ms}$\,m\,s$^{-2}$, where $P_{\rm ms}$ is the pulsar spin period in ms. Power from the first four harmonics was coherently summed to increase sensitivity to narrow pulses.
No pulsations were detected. The most constraining upper limit on the pulsed fraction was $<$26\% (99\% c.l.), based on the maximum Fourier power consistent with a non-detection across the full range of trial frequencies and accelerations.
Second, we performed a phase-modulation search on the whole \nustar\ observation using the \texttt{search}$_-$\texttt{bin} routine in \texttt{PRESTO}. This technique is designed to detect pulsar signals in compact binaries whose Fourier power is distributed over multiple orbital sidebands (for details, see \citealt{Ransom2003}). All candidates were subsequently folded with \texttt{prepfold}. Upon visual inspection, no promising signals were found.

We also performed a systematic search for X-ray bursts in the \nicer, \nustar, and EP/FXT data. We extracted time series using bin sizes of 1, 2, and 5\,s to probe bursts over a range of timescales. For each bin, we computed the Poisson probability of detecting the observed number of counts relative to the mean count rate, and converted this to a Gaussian-equivalent significance.
Bins exceeding a $5\sigma$ threshold were flagged as candidate bursts and subsequently cross-checked against background variations to rule out spurious events. Time intervals at the edges of Good Time Intervals (GTIs) were excluded, as abrupt changes in exposure at GTI boundaries can mimic short bursts. No statistically significant bursts were identified in any of the datasets.

\section{Periodicity search in ATLAS data}
\label{sec:periodsearch}

We searched for periodic variability in the combined ATLAS $o$- and $c$-band light curves collected between 2017 May 6 and 2024 October 22 using a Lomb–Scargle (LS) periodogram analysis \citep{Lomb1976, Scargle1982}. We computed the LS periodogram over the range 0.1--200\,d using a frequency grid tailored to the time baseline of the observations. To assess the impact of irregular sampling, we also computed the normalized spectral window function from the timestamps.

Figure\,\ref{fig:combined_periodogram} shows the LS periodogram, along with the corresponding spectral window function. To mitigate the risk of spurious detections arising from the temporal sampling pattern (see \citealt{Deeming1975, Vanderplas2018}), we rejected LS peaks whose frequencies were consistent with prominent features in the window function.
We then estimated false-alarm probability thresholds using a bootstrap resampling approach. No periodicity was detected above the 3$\sigma$ significance level.

\setcounter{figure}{0}
\begin{figure}[!h]
\centering
\includegraphics[width=0.47\textwidth]{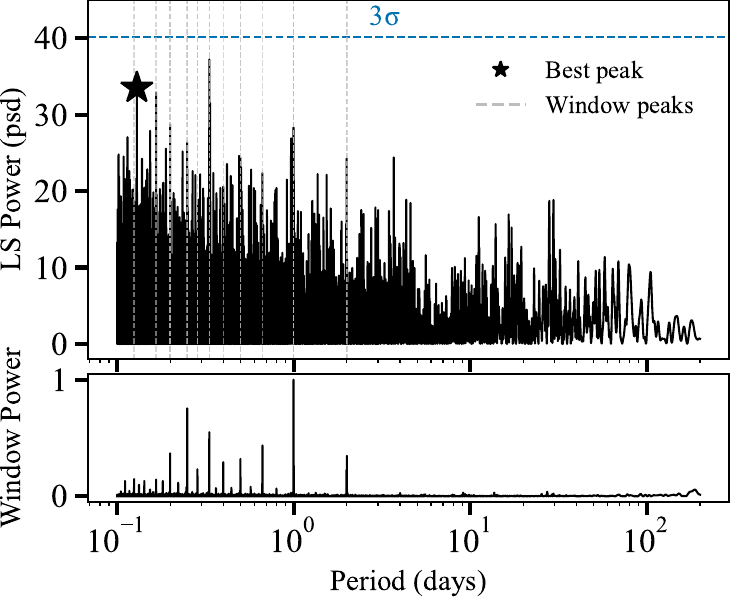}
\vspace{-0.3cm}
\caption{\emph{Top}: LS periodogram of the combined ATLAS $o$- and $c$-band light curve, computed over the range 0.1--200\,d. Vertical dashed lines mark the locations of prominent features in the spectral window function. The strongest non-alias peak is marked with a star. \emph{Bottom}: Normalized spectral window power, highlighting the alias structure introduced by the uneven sampling of the data.}
\label{fig:combined_periodogram}
\end{figure}

\section{Cross-Correlation Analysis of Optical and X-ray Light Curves}
\label{sec:zdcf}

To search for correlated optical and X-ray variability, we computed the Z-transformed Discrete Correlation Function (ZDCF; \citealt{Alexander1997,Alexander2013}) between the \nustar\ light curve and the light curves in the VT$_{\rm B}$- and VT$_{\rm R}$-bands from SVOM/VT, after converting VT$_{\rm B}$- and VT$_{\rm R}$-band magnitudes into fluxes.

Figure\,\ref{fig:zdcf} shows the resulting ZDCFs, computed over a lag range of $\pm$2000\,s. No statistically significant correlation peaks are found within the probed interval. We note that, due to the $\simeq$100\,s sampling of the VT optical light curves, potential correlations on shorter timescales remain inaccessible.

\setcounter{figure}{0}
\begin{figure}[!h]
\begin{center}
\includegraphics[width=0.47\textwidth]{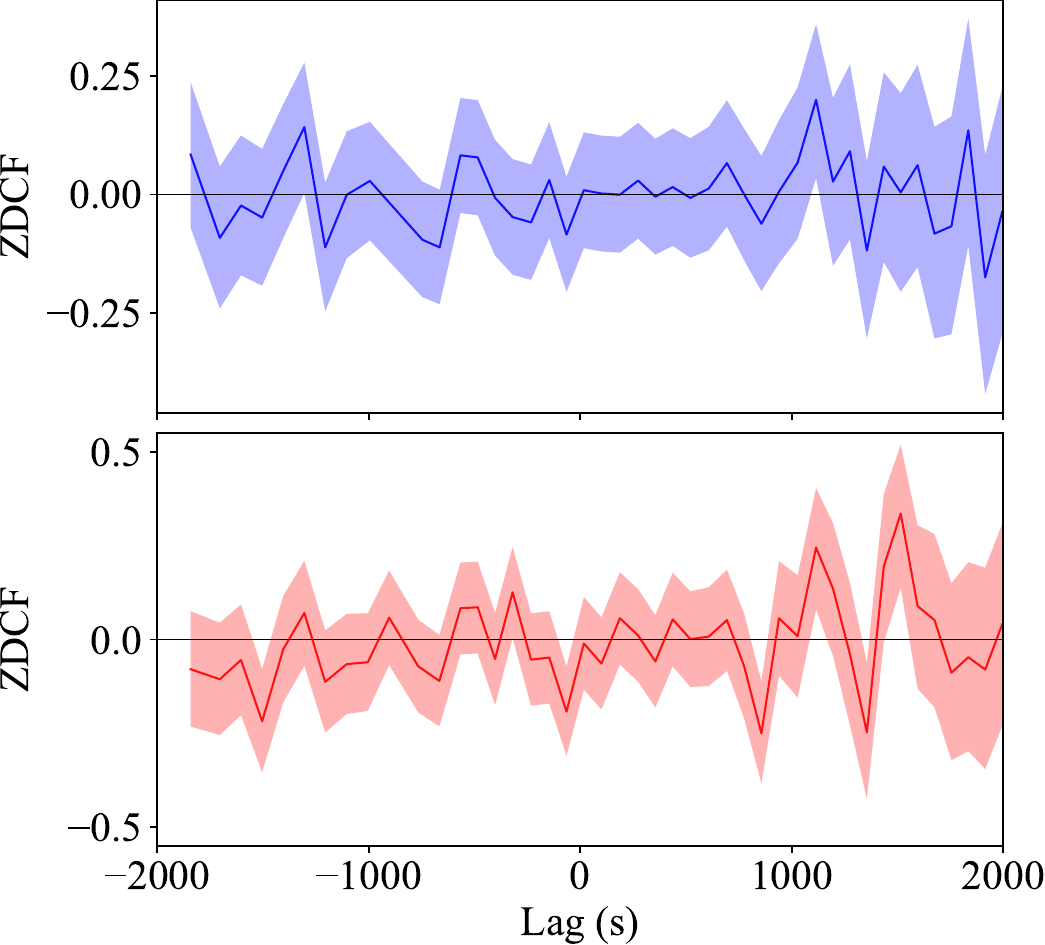}
\vspace{-0.5cm}
\caption{ZDCFs between the \nustar\ X-ray light curve and the SVOM/VT optical light curves in the VT$_{\rm B}$ (\emph{top}) and VT$_{\rm R}$ (\emph{bottom}) bands, computed over a lag range of $\pm$2000\,s. The solid lines represent the ZDCF values, while the shaded regions indicate the corresponding $1\sigma$ uncertainties. No statistically significant correlation peaks are detected.}
\label{fig:zdcf}
\end{center}
\end{figure}

\clearpage

\bibliography{biblio}
\bibliographystyle{aasjournalv7}



\end{document}